\documentclass[11pt,a4paper]{article}
\usepackage{jheppub}
\usepackage{appendix}
\usepackage{ifthen} 
\usepackage{overpic} %
\usepackage[utf8]{inputenc}
\usepackage[normalem]{ulem}
\newboolean{pdflatex}
\setboolean{pdflatex}{true} 

\newboolean{articletitles}
\setboolean{articletitles}{true} 

\newboolean{uprightparticles}
\setboolean{uprightparticles}{false} 
\usepackage{bm}

\usepackage{tabto} 
\usepackage{siunitx}
\usepackage{mathtools}
\usepackage{soul}
\DeclarePairedDelimiter\abs{\lvert}{\rvert}

\numberwithin{equation}{section}
\makeatletter
\let\oldtheequation\theequation
\renewcommand\tagform@[1]{\maketag@@@{\ignorespaces#1\unskip\@@italiccorr}}
\renewcommand\theequation{(\oldtheequation)}
\makeatother

\title{Maximising the physics potential of \texorpdfstring{\Bdpimumubold}{B->pimumu} decays}

\author[1]{Alexander Mclean Marshall,}
\author[2]{Michael Andrew McCann,}
\author[2]{Mitesh Patel,}
\author[1]{Konstantinos A. Petridis,}
\author[3,4]{M\'{e}ril Reboud,}
\author[3]{Danny van Dyk}

\affiliation[1]{H.H. Wills Physics Laboratory, University of Bristol, Bristol, BS8 1TL, UK}
\affiliation[2]{Blackett Laboratory, Imperial College London, London, SW7 2BW, UK}
\affiliation[3]{Institute for Particle Physics Phenomenology and Department of Physics, Durham University, Durham, DH1 3LE, UK}
\affiliation[4]{Physik Department, Universit\"{a}t Siegen, Walter-Flex-Str. 3, 57068 Siegen, Germany}

\emailAdd{alex.marshall@cern.ch}
\emailAdd{m.mccann@imperial.ac.uk}
\emailAdd{mitesh.patel@imperial.ac.uk}
\emailAdd{konstantinos.petridis@bristol.ac.uk}
\emailAdd{merilreboud@gmail.com}
\emailAdd{danny.van.dyk@gmail.com}

\abstract{\noindent We present a method that maximises the experimental sensitivity to new physics contributions in \BpmToPimm decays. This method relies on performing an unbinned maximum likelihood fit to both the measured dimuon \qsq distribution of \BpmToPimm decays, and theory calculations at spacelike \qsq, where QCD predictions are most reliable. Using known properties of the decay amplitude we employ a dispersion relation to describe the non-local hadronic contributions across spacelike and timelike \qsq regions. The fit stability and the sensitivity to new physics couplings and new sources of \CP-violation are studied for current and future data-taking scenarios, with the LHCb experiment as an example. 
The proposed method offers a precise and reliable way to search for new physics in these decays.

}

\usepackage{microtype}
\usepackage{lineno}  
\usepackage{xspace} 

\usepackage{graphicx}  
\usepackage{color}
\usepackage{colortbl}
\graphicspath{{./figs/}} 
\DeclareGraphicsExtensions{.pdf,.PDF,png,.PNG}

\usepackage{amsmath} 
\usepackage{amssymb}
\usepackage{amsfonts}
\usepackage{upgreek} 

\newcommand*\patchAmsMathEnvironmentForLineno[1]{%
\expandafter\let\csname old#1\expandafter\endcsname\csname #1\endcsname
\expandafter\let\csname oldend#1\expandafter\endcsname\csname
end#1\endcsname
 \renewenvironment{#1}%
   {\linenomath\csname old#1\endcsname}%
   {\csname oldend#1\endcsname\endlinenomath}%
}
\newcommand*\patchBothAmsMathEnvironmentsForLineno[1]{%
  \patchAmsMathEnvironmentForLineno{#1}%
  \patchAmsMathEnvironmentForLineno{#1*}%
}
\AtBeginDocument{%
\patchBothAmsMathEnvironmentsForLineno{equation}%
\patchBothAmsMathEnvironmentsForLineno{align}%
\patchBothAmsMathEnvironmentsForLineno{flalign}%
\patchBothAmsMathEnvironmentsForLineno{alignat}%
\patchBothAmsMathEnvironmentsForLineno{gather}%
\patchBothAmsMathEnvironmentsForLineno{multline}%
\patchBothAmsMathEnvironmentsForLineno{eqnarray}%
}

\usepackage{hyperxmp}

\usepackage[colorinlistoftodos,textsize=scriptsize]{todonotes}

\usepackage[all]{hypcap} 

\usepackage{xspace} 
\usepackage{upgreek}
\newcommand{\offsetoverline}[2][0.1em]{\kern #1\overline{\kern -#1 #2}}%

\def\qsq {\ensuremath{q^{2}}\xspace}

\ifthenelse{\boolean{uprightparticles}}%
{

 \def\Pmu         {\ensuremath{\upmu}\xspace}

 \def\Ppi         {\ensuremath{\uppi}\xspace}

 \def\PDelta      {\ensuremath{\Delta}\xspace}                 
 \def\PXi         {\ensuremath{\Xi}\xspace}                 
 \def\PLambda     {\ensuremath{\Lambda}\xspace}                 
 \def\PSigma      {\ensuremath{\Sigma}\xspace}                 
 \def\POmega      {\ensuremath{\Omega}\xspace}                 
 \def\PUpsilon    {\ensuremath{\Upsilon}\xspace}

 \def\PB      {\ensuremath{\mathrm{B}}\xspace}

 \def\PK      {\ensuremath{\mathrm{K}}\xspace}

 \def\Pi      {\ensuremath{\mathrm{i}}\xspace}

}
{

 \def\Pmu         {\ensuremath{\mu}\xspace}

 \def\Ppi         {\ensuremath{\pi}\xspace}

 \mathchardef\PDelta="7101
 \mathchardef\PXi="7104
 \mathchardef\PLambda="7103
 \mathchardef\PSigma="7106
 \mathchardef\POmega="710A
 \mathchardef\PUpsilon="7107
                  
 \def\PB      {\ensuremath{B}\xspace}

 \def\PK      {\ensuremath{K}\xspace}

 \def\Pi      {\ensuremath{i}\xspace}

}

\makeatletter
\ifcase \@ptsize \relax
  \newcommand{\miniscule}{\@setfontsize\miniscule{4}{5}}
\or
  \newcommand{\miniscule}{\@setfontsize\miniscule{5}{6}}
\or
  \newcommand{\miniscule}{\@setfontsize\miniscule{5}{6}}
\fi
\makeatother

\DeclareRobustCommand{\optbar}[1]{\shortstack{{\miniscule (\rule[.5ex]{1.25em}{.18mm})}
  \\ [-.7ex] $#1$}}

\def\mup        {{\ensuremath{\Pmu^+}}\xspace}
\def\mun        {{\ensuremath{\Pmu^-}}\xspace} 

\def\pion   {{\ensuremath{\Ppi}}\xspace}

\def\pipm   {{\ensuremath{\pion^\pm}}\xspace}

\def\kaon    {{\ensuremath{\PK}}\xspace}
  \def\Kbar    {{\kern 0.2em\overline{\kern -0.2em \PK}{}}\xspace}

\def\KorKbar {\kern 0.18em\optbar{\kern -0.18em K}{}\xspace}

\def\Kstarz  {{\ensuremath{\kaon^{*0}}}\xspace}

\def\B       {{\ensuremath{\PB}}\xspace}
\def\Bbar    {{\ensuremath{\kern 0.18em\overline{\kern -0.18em \PB}{}}}\xspace}

\def\BorBbar    {\kern 0.18em\optbar{\kern -0.18em B}{}\xspace}

\def\Bpm     {{\ensuremath{\B^\pm}}\xspace}

\def\Bd      {{\ensuremath{\B^0}}\xspace}

\def\Bs      {{\ensuremath{\B_{s}^0}}\xspace}

\newcommand{\decay}[2]{\mbox{\ensuremath{#1\!\to #2}}\xspace}         

\def\to                 {\ensuremath{\rightarrow}\xspace}

\def\qsq       {{\ensuremath{q^2}}\xspace}
\def\CP                {{\ensuremath{C\!P}}\xspace}

\def\BdToKstmm    {\decay{\Bd}{\Kstarz\mup\mun}}

\def\BupmToKmm      {\decay{\Bpm}{K^\pm\mu^+\mu^-}}

\def\BpmToPimm      {\decay{\Bpm}{\pi^\pm\mu^+\mu^-}}
\def\BpmToKmm      {\decay{\Bpm}{K^\pm\mu^+\mu^-}}

\def\BupmToKstmm    {\decay{\Bpm}{K^{*\pm}\mu^+\mu^-}}

\def\BsToPhimm    {\decay{\Bs}{\phi\mup\mun}}

\def\Bdpimumubold  {\boldmath{\decay{\Bpm}{\pipm\mu^+ \mu^-}}}

\def\C#1      {\ensuremath{\mathcal{C}_{#1}}\xspace}                       
\def\Cp#1     {\ensuremath{\mathcal{C}_{#1}^{'}}\xspace}                    
\def\Ceff#1   {\ensuremath{\mathcal{C}_{#1}^{\mathrm{(eff)}}}\xspace}        
\def\Cpeff#1  {\ensuremath{\mathcal{C}_{#1}^{'\mathrm{(eff)}}}\xspace}       
\def\Ope#1    {\ensuremath{\mathcal{O}_{#1}}\xspace}                       
\def\Opep#1   {\ensuremath{\mathcal{O}_{#1}^{'}}\xspace}                    

\newcommand{\aunit}[1]{\ensuremath{\text{\,#1}}}       

\newcommand{\tev}{\aunit{Te\kern -0.1em V}\xspace}
\newcommand{\gev}{\aunit{Ge\kern -0.1em V}\xspace}
\xspace
\newcommand{\mev}{\aunit{Me\kern -0.1em V}\xspace}
\newcommand{\kev}{\aunit{ke\kern -0.1em V}\xspace}
\newcommand{\ev}{\aunit{e\kern -0.1em V}\xspace}
\newcommand{\mevc}{\ensuremath{\aunit{Me\kern -0.1em V\!/}c}\xspace}
\newcommand{\gevc}{\ensuremath{\aunit{Ge\kern -0.1em V\!/}c}\xspace}
\newcommand{\mevcc}{\ensuremath{\aunit{Me\kern -0.1em V\!/}c^2}\xspace}
\newcommand{\gevcc}{\ensuremath{\aunit{Ge\kern -0.1em V\!/}c^2}\xspace}

\renewcommand{\gev}{\, \mathrm{GeV}}
\renewcommand{\tev}{\, \mathrm{TeV}}

\renewcommand{\C}[1]{{\cal C}_{#1}}
\renewcommand{\Cp}[1]{{\cal C}_{#1'}}

\newcommand{\Cnineeff}{\ensuremath{C_9^\mathrm{eff, B^\pm}}\xspace}

\usepackage{mciteplus}
\usepackage{multirow}
\usepackage{booktabs}

\begin{document}

\newcolumntype{L}[1]{>{\raggedright\let\newline\\\arraybackslash\hspace{0pt}}m{#1}}
\newcolumntype{C}[1]{>{\centering\let\newline\\\arraybackslash\hspace{0pt}}m{#1}}
\newcolumntype{R}[1]{>{\raggedleft\let\newline\\\arraybackslash\hspace{0pt}}m{#1}}

\renewcommand{\thefootnote}{\fnsymbol{footnote}}
\setcounter{footnote}{1}

\renewcommand{\thefootnote}{\arabic{footnote}}
\setcounter{footnote}{0}
\maketitle
\flushbottom

\cleardoublepage

\pagestyle{plain} 
\setcounter{page}{1}
\pagenumbering{arabic}


\section{Introduction}

    Over the past decade, several experimental results have hinted at the possibility of new physics in $b\to s\ell^+\ell^-$ transitions. Most notably, deviations from the predictions of the Standard Model have been observed in the decay rates of \BdToKstmm, \BupmToKstmm, \BupmToKmm, and \BsToPhimm decays~\cite{LHCb:2015wdu,LHCb:2016ykl,LHCb:2021zwz,BELLE:2019xld,Belle:2019oag,Belle:2016fev}; and angular distributions of \BdToKstmm, \BupmToKstmm and \BsToPhimm transitions~\cite{LHCb:2020gog,LHCb:2020lmf,LHCb:2021xxq,ATLAS:2018gqc,CMS:2020oqb,CMS:2015bcy}. The signs of electron-muon universality violation in  $b\to s\ell^+\ell^-$ have all but evaporated as presented by the recent updates to $R_{K}$ and $R_{K^*}$ measurements by the LHCb collaboration~\cite{LHCb:2022vje}. This suggests that the decay rates and angular distributions of $b\to s e^+e^-$ processes exhibit the same tensions with SM predictions as their muon counterparts. 
    
    Global analyses of these updated measurements point predominantly to anomalous couplings between a left-handed $\bar{s}b$ current and a vectorial lepton current~\cite{Alguero:2023jeh,Ciuchini:2022wbq}. Such a hint is quantitatively supported by, separately, branching ratios and angular $b \to s$ data \cite{Guadagnoli:2023ddc}, whose systematic uncertainties are generally very different.
    A more mundane explanation of the experimental measurements involves underestimating hadronic contributions in the SM~\cite{Ciuchini:2022wbq}. Such hadronic effects involve non-local matrix elements of four-quark operators that are hard to compute from first principles. However, recent re-appraisals of these hadronic components suggest they are less likely to be the cause of the observed anomalies in $b\to s\mu^+\mu^-$ decays~\cite{Gubernari:2022hxn}. 
    Such a conclusion could be validated by suitable observables at high $q^2$, which share the very same short-distance sensitivity while not suffering from the same long-distance issues \cite{Guadagnoli:2023ddc}. These observables include $B_s \to \mu^+ \mu^- \gamma$ \cite{Dettori:2016zff} and the inclusive $B \to Xs \mu^+ \mu^-$ \cite{Isidori:2023unk} among the others.
    
    Traditionally, measurements of  $b\to s\mu^+\mu^-$ transitions involve binning the data in regions of the invariant mass of the dimuon system squared ($\qsq$) and performing measurements of decay rates and angular observables within each of these bins. Recent developments in theory and experiment have opened up the possibility of fitting the entirety of the differential decay rate of $B\to K^{(*)}\mu^+\mu^-$ transitions to determine new physics couplings and hadronic contributions from the data~\cite{Blake:2017fyh,Chrzaszcz:2018yza,Cornella:2020aoq}.

    The additional CKM (Cabibbo–Kobayashi–Maskawa matrix) suppression in the SM of $b\to d\ell^+\ell^-$ relative to $b\to s\ell^+\ell^-$ processes makes observables of the former even more sensitive probes of new physics~\cite{Biswas:2022lhu}. In light of the tensions with SM predictions in $b\to s \ell^+\ell^-$ processes, maximising the experimental sensitivity in \BpmToPimm decays is of paramount importance to ascertain a more complete picture of the flavour structure of these tensions, be they due to new physics or hadronic effects. Recently, branching fraction measurements of \BpmToPimm~\cite{LHCb:2015hsa}, $B_{s}^{0}\to \bar{K}^{*0}\mu^+\mu^-$~\cite{LHCb:2018rym} and $B^0\to\mu^+\mu^-$~\cite{LHCb:2021awg,CMS:2022mgd} decays have been combined to constrain new physics contributions in $b\to d\ell^+\ell^-$ processes~\cite{Rusov:2019ixr,Bause:2022rrs}. However, such analyses suffer from limited experimental precision and coarse information regarding the $\qsq$ distribution of \BpmToPimm processes. The analysis of Ref.~\cite{Bordone:2021olx} uses a dispersive model for the non-local contributions in $b\to d\ell^+\ell^-$ transitions to predict lepton flavour universality ratios, for which hadronic uncertainties largely cancel. However, in order to ascertain new physics contributions in lepton-flavour-specific final states, it is imperative to separate long- and short-distance effects. This can only be done through an unbinned fit to the dimuon spectrum of $B^\pm\to\pi^\pm\mu^+\mu^-$ transitions adopting an effective field theory description of the decay amplitudes. Additionally, as will be demonstrated, employing QCD factorisation and light-cone sum rules (LCSR) predictions at negative \qsq to constrain the size of hadronic contributions is essential to maximise sensitivity to new physics in these decays, the incorporation of this information is the primary innovation of this paper.

    This paper is organised as follows: \autoref{sec:Theoretical_Framework} introduces the theoretical background and provides a description of the model used, \autoref{sec:Fit_Description} describes how the fits to pseudo-datasets are set up, \autoref{sec:experimental_prospects_and_precision} details our results and finally \autoref{sec:conclusion} provides a conclusion. 
    
\section{Theoretical Framework}
\label{sec:Theoretical_Framework}

    We work within the usual weak effective theory for low-energy $b\to d\ell^+\ell^-$ transitions. Its effective Lagrangian reads~\cite{Bobeth:1999mk,Bobeth:2001jm}
    \begin{equation}
        \label{eq:th:Leff}
        \mathcal{L}_\mathrm{eff}^{bd\ell\ell}
            = \frac{4 G_F}{\sqrt{2}} \left(\lambda_c \, \mathcal{L}_\mathrm{eff}^{(c)} + \lambda_u \, \mathcal{L}_\mathrm{eff}^{(u)} \right)
            + \text{h.c.},
    \end{equation}
    where we abbreviate the CKM factors $\lambda_q = V_{qb}^{\phantom{*}} V_{qd}^*$ and use
    \begin{equation}
        \label{eq:th:Leff-p}
        \mathcal{L}_\mathrm{eff}^{(p)} = \mathcal{C}_1 \mathcal{O}_1^p + \mathcal{C}_2 \mathcal{O}_2^p
            + \sum_{i \in \mathcal{I}} \mathcal{C}_i \mathcal{O}_i.
    \end{equation}
    Above, the sums run over the set of operators $\mathcal{I} = \{3 - 10, 7' - 10', P, P', S, S', T, T5\}$. These operators are commonly classified as either semileptonic ($9,9',10,10',P,P',S,S',T,T5$), radiative ($7,7',8,8'$), current-current ($1,2$) and QCD penguin operators ($3 - 6$). In contrast to $b\to s$ transitions, $b\to d$ transitions exhibit a flat hierarchy of the CKM factors $\lambda_u \sim \lambda_c \sim \lambda_t$, which requires one to keep all the terms in \autoref{eq:th:Leff} in the calculations. Note that we allow for BSM physics to enter the weak effective theory through the semileptonic operators $\mathcal{O}_i$ with $i=9,9',10,10',P,P',S,S',T,T5$ only. This procedure follows what is done in $b\to s\ell^+\ell^-$ transitions.

\begin{figure}[t]
        \centering
        \includegraphics[width=0.32\textwidth]{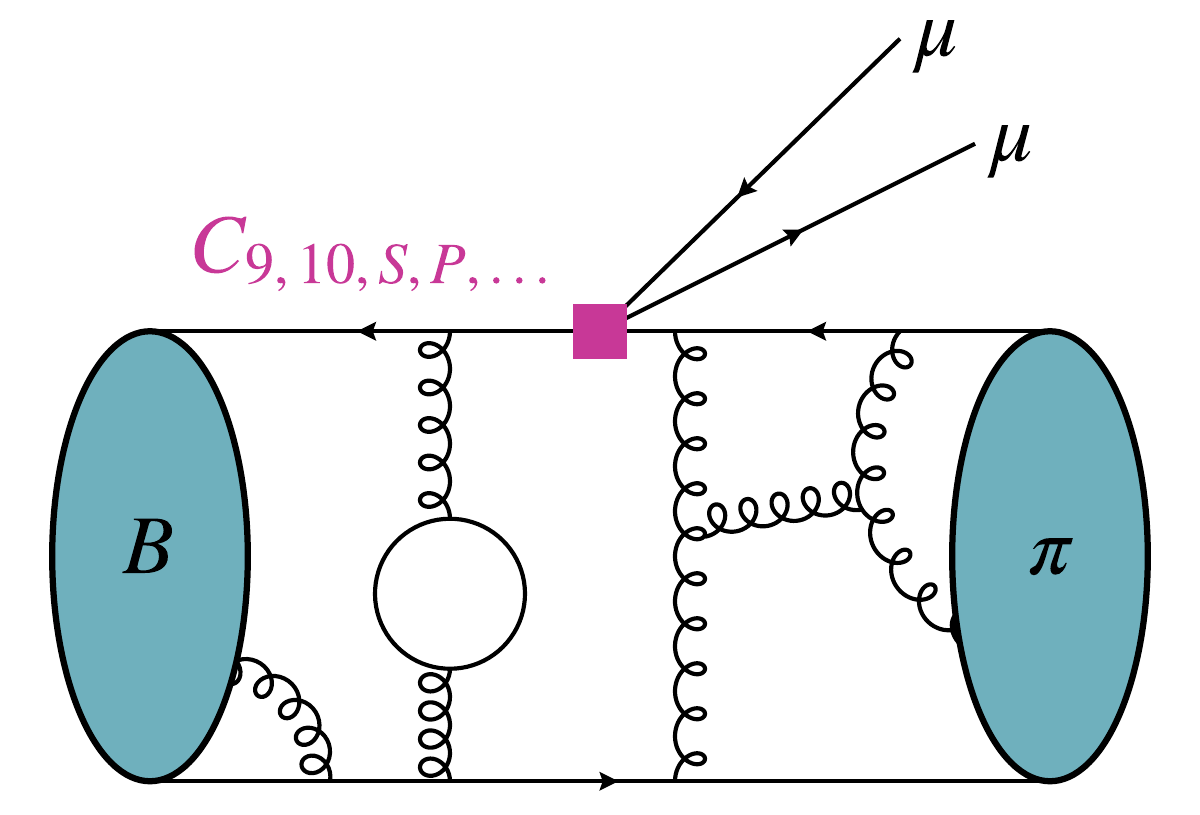}
        \includegraphics[width=0.32\textwidth]{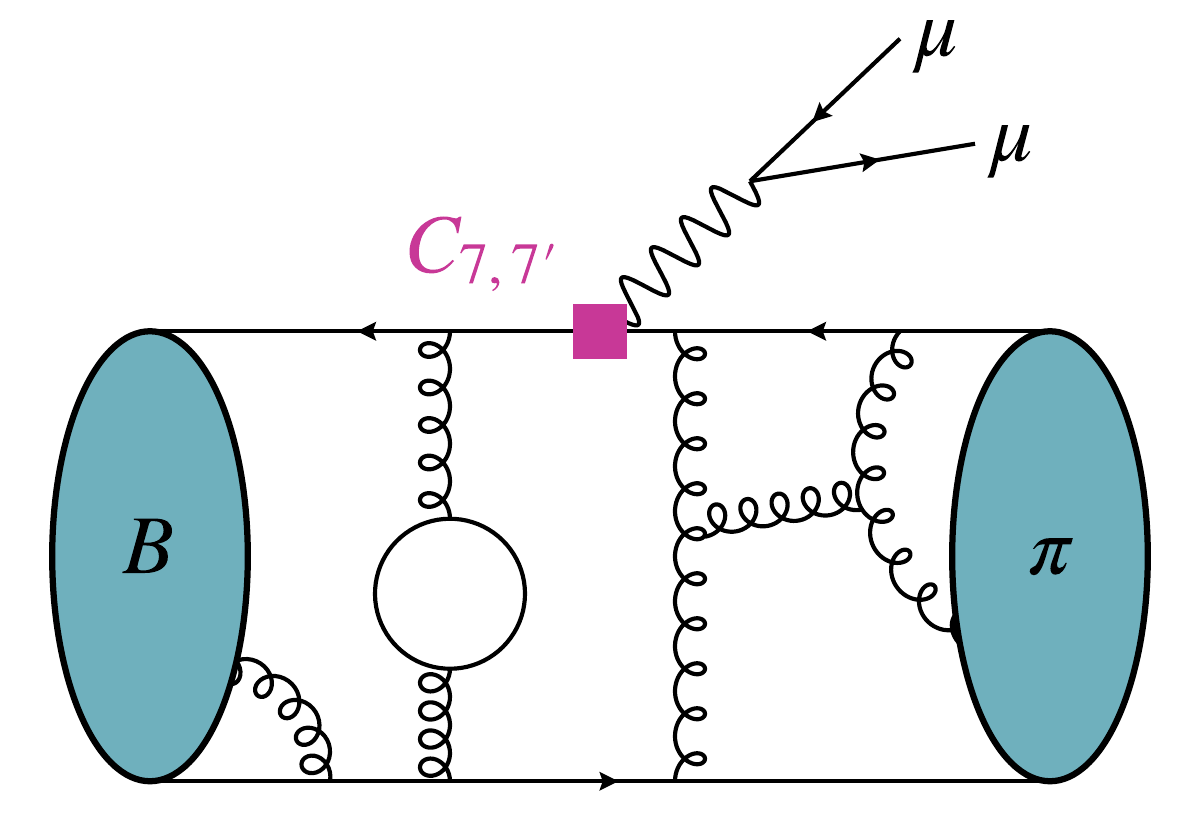}
        \includegraphics[width=0.32\textwidth]{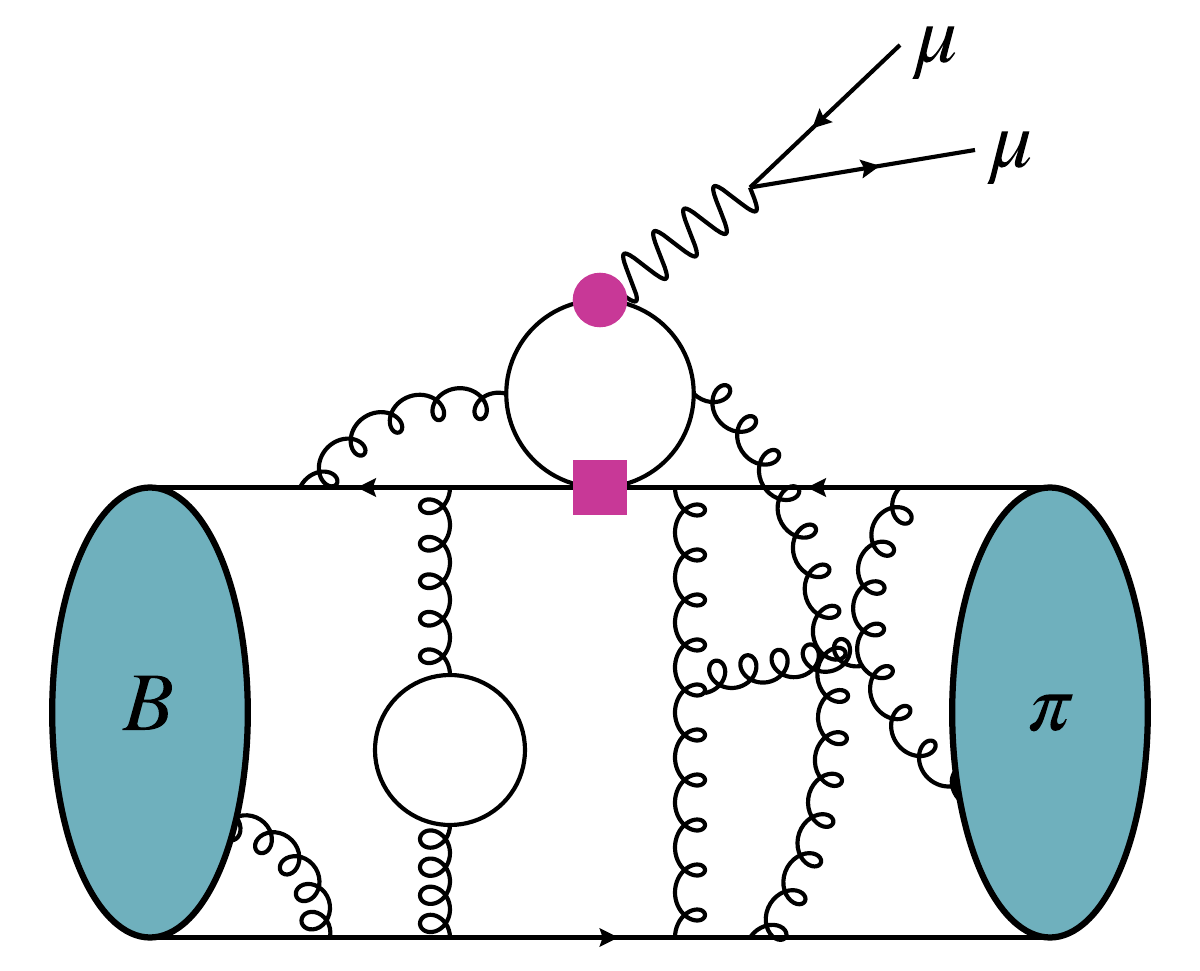}
        \caption{%
            Schematic overview of the two classes of contributions to $B^\pm\to\pi^\pm\mu^+\mu^-$ decays.
            The local contributions are presented in the left and central sketches,
            and one example of the non-local contributions is presented in the right sketch.}
        \label{fig:theory:illustration}
    \end{figure}

    The matrix elements arising from these effective operators can be classified as either local form factors or non-local form factors. Local form factors enter the amplitudes through the hadronic matrix elements of a two-parton current, e.g., from the semileptonic operators or the QED radiative operators $i=7,7'$. Non-local form factors enter the amplitudes through the time-ordered product of the electromagnetic current with effective operators: the four-quark current-current or QCD penguin operators; and radiative operators with $i=8,8'$. In \autoref{fig:theory:illustration} we provide a schematic overview of the two classes of contributions.
    
\begin{figure}[t]
        \centering
        \includegraphics[width=0.6\textwidth,keepaspectratio]{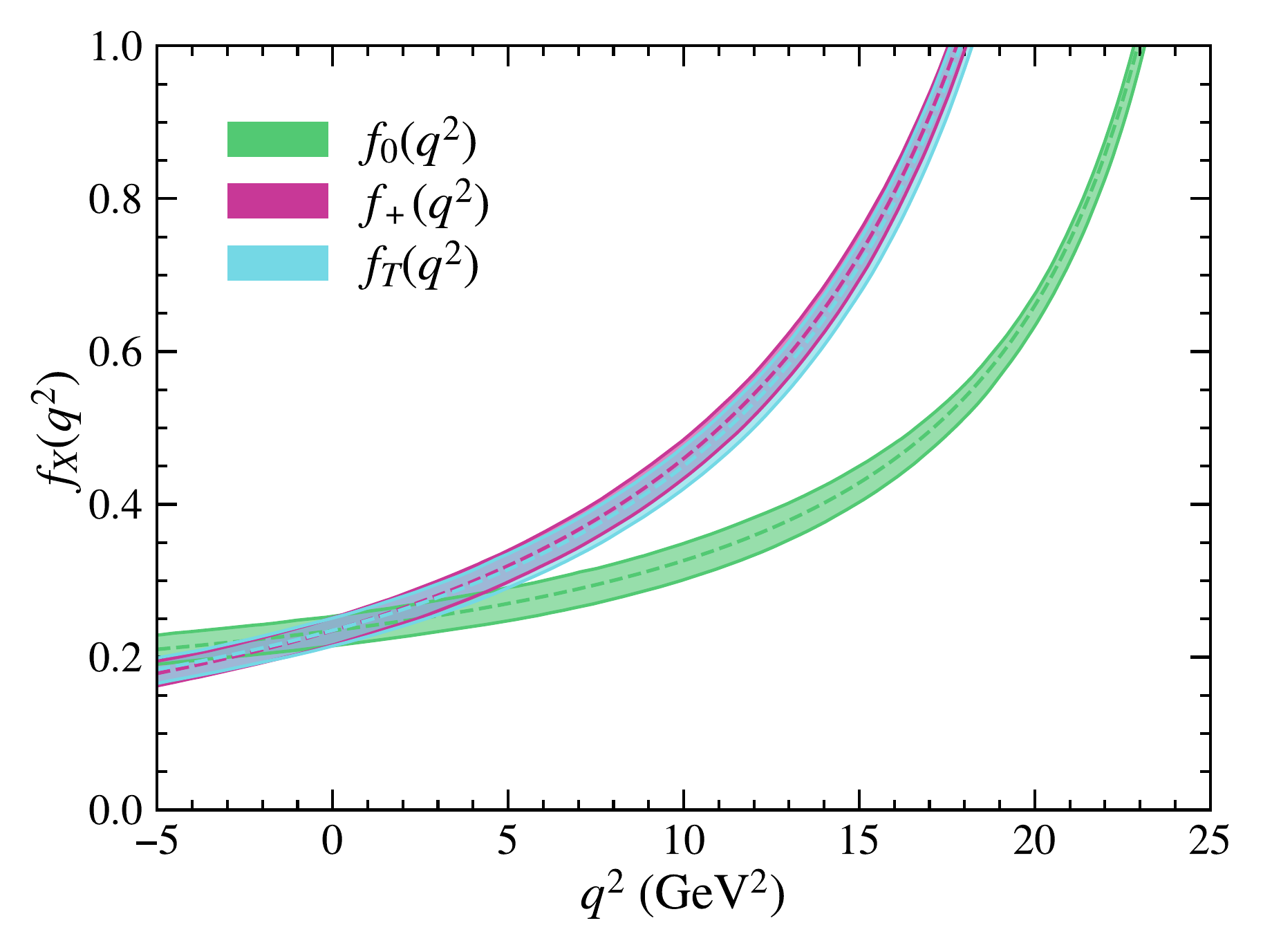}
        \caption{$\bar{B}\to \pi$ local form factors obtained in Ref.~\cite{Leljak:2021vte} by a combined fit to lattice QCD and light-cone sum rule estimates.
            The bands correspond to the $68\%$ interval.}
        \label{fig:FF_plots}
    \end{figure}

    In the case of $\bar{B}\to \pi$ transitions, there exist only three local form factors,
    which are labelled $f_+, f_0$ and $f_T$. Other form factors must vanish due to Lorentz invariance
    and parity conservation within the strong interaction. The three form factors are defined \textit{via}:
    \begin{align}
        \langle\bar \pi(k) | \bar b \gamma^\mu d | \bar{B}(p) \rangle &=
            \left[ (p + k)^\mu - \frac{M_B^2 - M_\pi^2}{q^2} q^\mu \right] f_+(q^2)
            + \frac{M_B^2 - M_\pi^2}{q^2} q^\mu f_0(q^2), \\[3mm]
        \langle\bar  \pi(k) | \bar b \sigma^{\mu\nu} q_\nu d | \bar{B}(p) \rangle &=
            \frac{i}{M_B + M_\pi} \left[ q^2 (p + k)^\mu - (M_B^2 - M_\pi^2) q^\mu \right] f_T(q^2).
    \end{align}
    The form factors are scalar-valued functions of the momentum transfer $q^2$, which requires
    some form of parametrization. Here, we use the nominal parametrization and numerical results from Ref.~\cite{Leljak:2021vte}. The parametrization used is based on the original BCL parametrization~\cite{Bourrely:2008za}. The numerical results are obtained from a combined fit to lattice QCD~\cite{FermilabLattice:2015mwy,FermilabLattice:2015cdh,Flynn:2015mha} and LCSR~\cite{Ball:2004ye,Duplancic:2008ix,Leljak:2021vte} inputs. We display these form factor results in \autoref{fig:FF_plots}.
    
    The Lagrangian density \autoref{eq:th:Leff} gives further rise to non-local contributions, stemming either from the full set of four-quark operators or the radiative operators with $i=8,8'$. In the case of $\bar{B}\to \pi$ transitions, there exists only a single Lorentz structure for these non-local contributions:
    \begin{align}
        \mathcal{H}^{(p),B^\pm}_\mu &= i \int d^4x \, e^{i q\cdot x}
            \langle \pi(k) | \mathcal{T}\left\{ j_\mu^\mathrm{em}(x), \mathcal{C}_1 \mathcal{O}_1^p(0) + \mathcal{C}_2 \mathcal{O}_2^p(0) + \sum_{i\,\in\,\{3-6,8,8'\}} \mathcal{C}_i \mathcal{O}_i(0) \right\}| B^\pm(p) \rangle \nonumber \\
        & = -\frac{1}{2} \left[ q^2 (p + k)^\mu - (M_B^2 - M_\pi^2) q^\mu \right] \mathcal{H}^{(p), B^\pm}(q^2),  \quad(p=u, c).
    \end{align}
    The non-local contributions can be recast into a shift to the Wilson coefficient $C_9$ via:
    \begin{equation}
    \begin{split}
        \Delta C_9^{B^+}(q^2) = -16\pi^2 \, \frac{\lambda_u \mathcal{H}^{(u),B^+}(q^2) + \lambda_c \mathcal{H}^{(c),B^+}(q^2)}{\lambda_t f^+(q^2)}, \\
        \Delta C_9^{B^-}(q^2) = -16\pi^2 \, \frac{\lambda_u^* \mathcal{H}^{(u),B^-}(q^2) + \lambda_c^* \mathcal{H}^{(c),B^-}(q^2)}{\lambda_t^* f^+(q^2)}.
    \end{split}
        \label{deltaC9_relation}
    \end{equation}
    Due to the \CP-violating nature of the weak interaction, we must take care to define such a shift separately for the $B^+$ and the $B^-$ initial state.

    Using the above definitions, the differential decay rate for the \BpmToPimm reads \cite{Ali:2013zfa}, 
    \begin{equation} \label{eq1}
    \begin{split}
    \frac{d\Gamma(\BpmToPimm)}{dq^2} & =  \frac{G^2_F\alpha^2|V_{tb}^{\phantom{*}}V^*_{td}|^2}{2^7\pi^5}|k|\biggl\{\frac{2}{3}|k|^2\beta_+^2|C_{10}f_+(q^2)|^2 \\
    &+ \frac{m_\mu^2(M^2_B-M^2_\pi)^2}{q^2M^2_B}|C_{10}f_0(q^2)|^2 \\
    &+ |k|^2 \biggr[1-\frac{1}{3}\beta^2_+\biggr]\bigg|\Cnineeff(q^2)f_+(q^2)+2C_7\frac{m_b+m_d}{M_B+M_\pi}f_T(q^2)\bigg|^2\biggl\},
    \end{split}
    \end{equation}
    where $q^2=m_{\mu\mu}^2$ and $|k|=\sqrt{E_\pi^2-M_\pi^2}$. This decay rate is defined separately for $B^+\rightarrow\pi^+\mu^+\mu^-$ and $B^-\rightarrow\pi^-\mu^+\mu^-$, with each having a unique \Cnineeff term defined as follows,
    \begin{equation}
        \Cnineeff(q^2) = |C_9|e^{\pm i\delta_{C_9}} + \Delta C_{9}^{B^\pm}(q^2).
    \end{equation}
    
    \subsection{Modelling the non-local contributions}
    \label{sec:hadronic}
    
        In the $q^2<0$ region, it is possible to compute the size of the non-local contributions to \BpmToPimm transitions using QCD factorization and LCSR. The individual non-local components are labelled as follows: factorizable loops $\mathcal{H}_{\text{fact,LO}}^{(p)}$, weak annihilation $\mathcal{H}_{\mathrm{WA}}^{(p)}$, factorizable NLO contributions $\mathcal{H}_{\text {fact,NLO}}^{(p)}$, non-factorizable soft-gluon contributions $\mathcal{H}_{\mathrm{soft}}^{(p)}$ and $\mathcal{H}_{\mathrm{soft}, \mathrm{O}_8}^{(p)}$, and non-factorizable spectator scattering $\mathcal{H}_{\text{nonf,spect}}^{(p)}$, where the $B^\pm$ index is dropped for legibility. The individual components are provided in Sec.~3 of Ref.~\cite{Hambrock:2015wka}. These components are summed to compute the full non-local contribution as in the following expression,
        \begin{equation}
            \begin{split} \label{eq:H_sum}
             \mathcal{H}^{(p)}\left(q^2\right) = \mathcal{H}_{\text{fact,LO}}^{(p)}\left(q^2\right)
            & + \mathcal{H}_{\mathrm{WA}}^{(p)}\left(q^2\right)
            + \mathcal{H}_{\text{fact,NLO}}^{(p)}\left(q^2\right)
            + \mathcal{H}_{\mathrm{soft}}^{(p)}\left(q^2\right) \\
            & + \mathcal{H}_{\mathrm{soft}, \mathcal{O}_8}^{(p)}\left(q^2\right)
            + \mathcal{H}_{\text{nonf,spect}}^{(p)}\left(q^2\right).
            \end{split}
        \end{equation}
        To model $\mathcal{H}^{(p), B^\pm}\left(q^2\right)$ across the full $q^2$ range, including the physical $q^2>0$ region, we employ once-subtracted dispersion relations\footnote{We refer to our model as a once-subtracted dispersion relation
following the nomenclature used in Refs.~\cite{Hambrock:2015wka,
Cornella:2020aoq}. However, while our model is inspired by a dispersion
relation it does not qualify as one on mathematical grounds, as
discussed later in the section.}
as in Eq.~(41) of Ref.~\cite{Hambrock:2015wka}. Combining with \autoref{deltaC9_relation} results in the following relation,
        \begin{equation}
            \Delta C_9^{B^\pm}(q^2) = \Delta C_9^{B^\pm}(q_0^2)
             + Y_{\rho, \omega}^{B^\pm}(q^2) + Y^{B^\pm}_\mathrm{LQC}(q^2) 
             + Y_{J/\psi, \psi(2S), ...}^{B^\pm}(q^2) + Y^{B^\pm}_{\text{2P}, c\bar{c}}(q^2).\label{eq:dispersion}
        \end{equation}
        To ensure the convergence of the dispersive integral for $\mathcal{H}^{(p), B^\pm}(q^2)$, we require one subtraction in the dispersion relation. The emerging subtraction terms are matched to the results of the QCD factorisation and LCSR calculations at the subtraction point $q^2_0$, as originally proposed in Ref.~\cite{Hambrock:2015wka}. For this analysis, we choose the subtraction point $q^2_0=-1.5 ~\mathrm{GeV}^2$. Finally, the various $Y^{B^\pm}(q^2)$ terms are the individual components of the non-local contributions that will be introduced in the following paragraphs.
    
        \paragraph{Resonances}
        \label{sec:res}
        
            The resonances considered within the full $q^2$ spectra of \BpmToPimm decays are the $\rho(770)$, $\omega(782)$, $J/\psi$, $\psi(2S)$, $\psi(3770)$, $\psi(4040)$, $\psi(4160)$, and the $\psi(4415)$. As in Ref.~\cite{Hambrock:2015wka}, we ignore the presence of the $\phi(1020)$ since its production is either OZI suppressed (in the production through current-current operators) or suppressed by small values of the SM Wilson coefficients (in the production through QCD penguin operators).
            
            Each resonance ($V$) contribution to $\Delta C_9^{B^\pm}(q^2)$ is described with a relativistic Breit-Wigner distribution as follows,
            \begin{equation}
                Y_{V}^{B^\pm}(q^2) = \eta^{B^\pm}_V e^{i\delta^{B^\pm}_V} \frac{(q^2-q^2_0)}{(m^2_V-q^2_0)}\frac{m_V\Gamma_{0V}}{(m^2_V-q^2)-im_V\Gamma_V(q^2)}.
            \end{equation}
            Here $\eta_V$ is the resonance magnitude, $\delta_V$ its phase\footnote{%
                Contrary to what is done in the description of exclusive $b\to s\mu^+\mu^-$ decays, the phases in our hadronic model for the non-local contributions are not strong phases; instead, they are superpositions of two strong phases arising from the two terms and the relative weak phase in \autoref{eq:th:Leff}.
            }, and $\Gamma_V(q^2)$ the running width, 
            \begin{equation}
            \begin{split}
                \Gamma_V(q^2) = \frac{p(q^2)}{p(m_V^2)}\frac{m_V}{\sqrt{q^2}}\Gamma_{0V},
                \quad\text{where}\quad
                p(q^2) = \frac{\sqrt{\lambda(q^2, m_\mu^2, m_\mu^2)}}{2\sqrt{q^2}}, \\
                \quad\text{where}\quad \lambda(A, B, C) = A^2 + B^2 + C^2 - 2(AB + AC + BC).
            \end{split}
            \end{equation}
            The description of the width involves the breakup momentum $p$ both as a function of $q^2$ and evaluated at $q^2 = m_{V}^2$. Our choice of the description of the residues in terms of two magnitude and phases, one each for the $B^+$ and $B^-$ decay, facilitates the description of \CP-violation in the decay.
            
        \paragraph{Open charm continuum}
        \label{sec:DDbar}

            We jointly model the combination of the non-resonant continuum of open charm states and the contributions due to further broad vector charmonia following the model suggested in Ref.~\cite{Cornella:2020aoq}. This model is governed by an overall coupling strength for the modelled two-particle open charm continuum and further includes terms for the $S$- and $P$-wave contributions. As for the resonance terms, we choose to describe each coupling in terms of a magnitude $\eta$ and a phase $\delta$, to facilitate the description of \CP-violation in the decay. In contrast to our modelling of the resonances, we choose to use the same coupling strength for both $B^+$ and $B^-$ decay. The model expression reads:
            \begin{equation} 
            \begin{split}
                 Y^{B^+}_{2 \mathrm{P},c \bar{c}}\left(q^{2}\right)
                 &=\eta_{2 \mathrm{P}}e^{i\delta_{2 \mathrm{P}}}\sum_{j=D^*D, D^*D^*, DD} \eta_j e^{i\delta_j} \frac{(q^2-q^2_0)}{\pi}\int_{s_0^j}^\infty \frac{d s}{(s-q^2_0)}\frac{
                 \hat \rho_j (s)}{(s-q^2)}\,, \\
                 Y^{B^-}_{2 \mathrm{P},c \bar{c}}\left(q^{2}\right) &= Y^{B^+}_{2 \mathrm{P},c \bar{c}}\left(q^{2}\right),
                 \label{eq:charm2p}
            \end{split} 
            \end{equation} 
            where $\hat{\rho}_i$ are hadronic spectral densities defined in Ref.~\cite{Cornella:2020aoq} and we use the same subtraction point $q^2_0=-1.5 ~\mathrm{GeV}^2$ as before. We fix the magnitudes $\eta_{D^*D}$, $\eta_{D^*D^*}$ and $\eta_{DD}$ of the modelled contributions to unity and fix the phases $\delta_{D^*D}$, $\delta_{D^*D^*}$ and $\delta_{DD}$ to zero. In contrast, the ``global'' parameters $\eta_{2 \mathrm{P}}$ and $\delta_{2 \mathrm{P}}$ are allowed to vary in fits to pseudo-data.

            The joint modelling of the heavy charmonium resonances as one-body intermediate states and the two-particle continuum amplitudes inevitably leads to some double counting and model error. We expect this to be insignificant compared to the statistical uncertainties achievable with the upcoming LHCb datasets. To validate this assumption, we assess the impact of this model choice on the measurement of the Wilson coefficients
            $\mathcal{C}_9$ and $\mathcal{C}_{10}$. We perform fits to pseudo-data generated with the default non-local amplitudes, including the ones above the open charm threshold, and fit back with variations of the non-local amplitude that involve turning off individual open-charm resonant and two-particle amplitudes. The resulting variations on the extracted values of $\mathcal{C}_9$ and $\mathcal{C}_{10}$ are found to be negligible compared to the statistical precision of any current or future experiment.

        \paragraph{Light-quark continuum}\label{LQC}

            Finally, we need to consider the non-local contribution from the ``light-quark'' continuum, i.e., the continuum of $\bar{u}u$, $\bar{d}d$ and $\bar{s}s$ states. In a perturbative picture, this contribution arises from weak annihilation and light-quark loop diagrams. This contribution is modelled using the following integral over hadronic spectral densities,
            \begin{equation}
            Y^{B^\pm}_\mathrm{LQC}(q^2) =  \sum_{q=u,c} \int_{s_0\;\simeq\;1.5\;\mathrm{GeV}^2}^{4m_D^2} ds\frac{(q^2-q^2_0)\rho^{(q^\pm)}_\mathrm{LO}(s)}{(s-q^2_0)(s-q^2-i\sqrt{s}\Gamma_{\mathrm{eff}}(s))},
            \end{equation}
            where $\rho^{(u)}_\mathrm{LO}(s)$, $\rho^{(c)}_\mathrm{LO}(s)$ and $\Gamma_{\mathrm{eff}}$ are provided in Eq.~38, Eq.~39 and in the text of Ref.~\cite{Hambrock:2015wka}, respectively. Using a duality threshold $s_0=1.5$\;GeV$^2$ reduces the impact of any potential double counting between the $\rho(770)$ and the $\omega(782)$ and the light-quark continuum. The physical quantities that build up this component are known well enough such that $Y^{B^\pm}_\mathrm{LQC}(q^2)$ is fixed in the fit.

        \begin{figure}[t]
        \centering
        
        \includegraphics[width=0.48\textwidth,keepaspectratio]{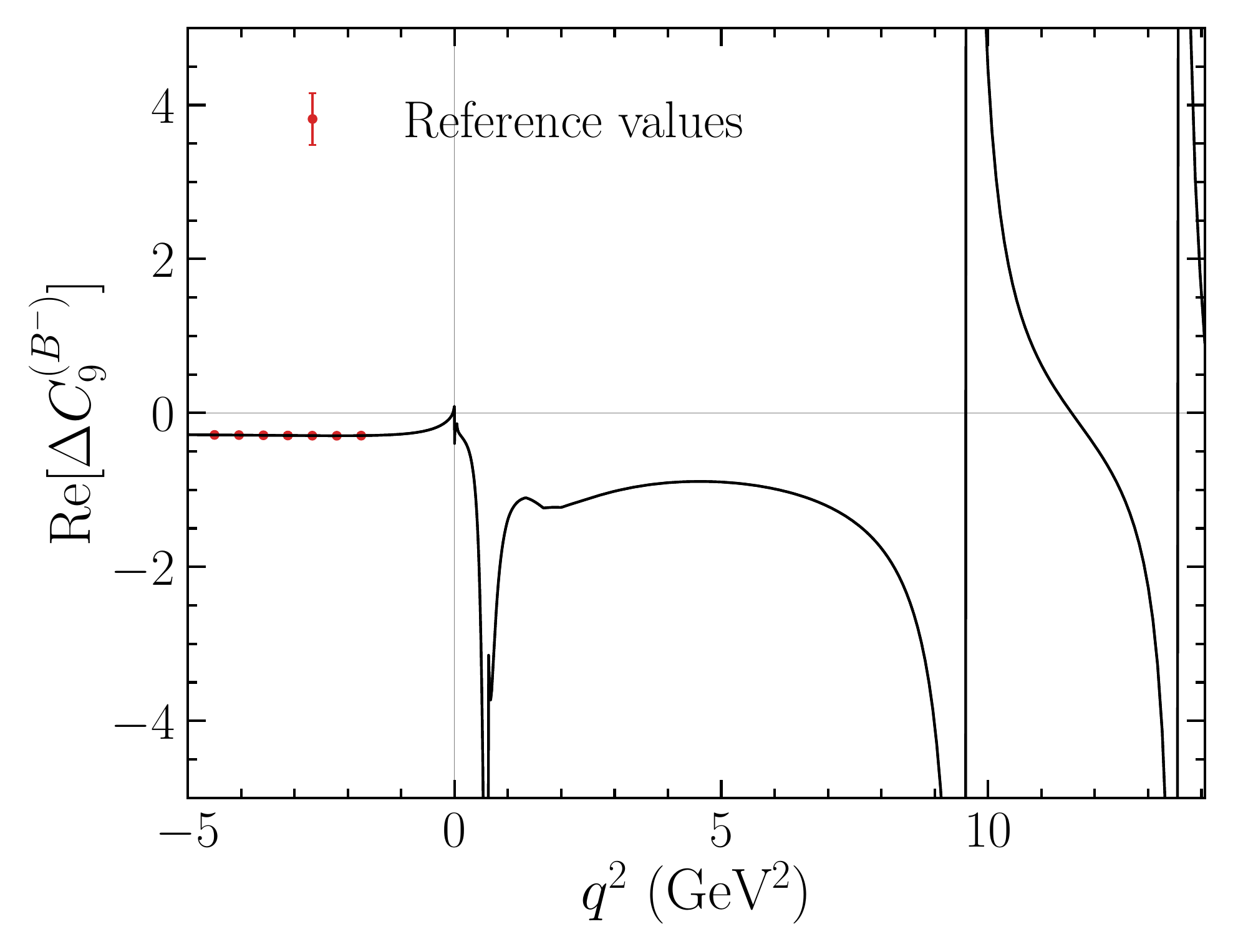}
        \includegraphics[width=0.48\textwidth,keepaspectratio]{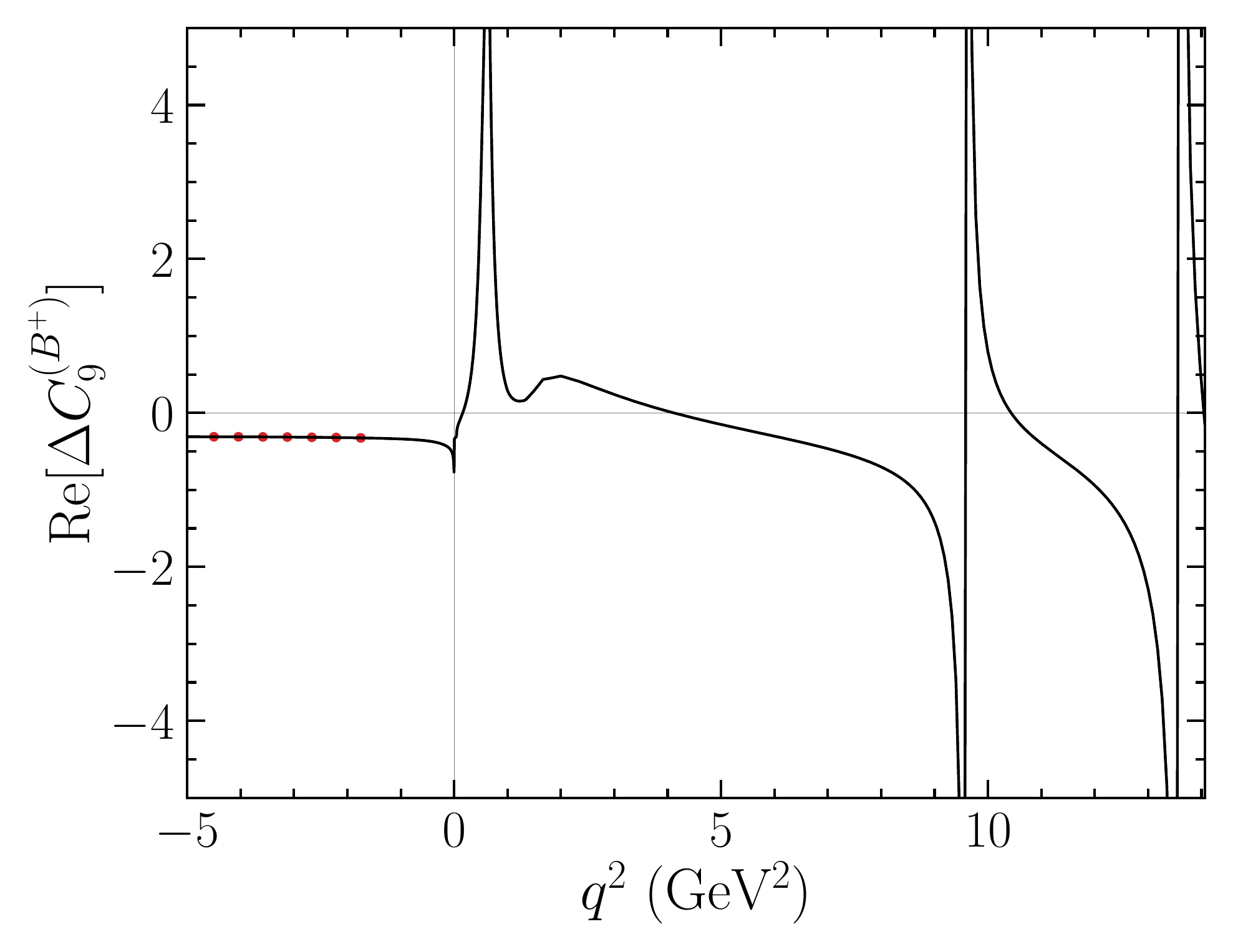}
        
        \includegraphics[width=0.48\textwidth,keepaspectratio]{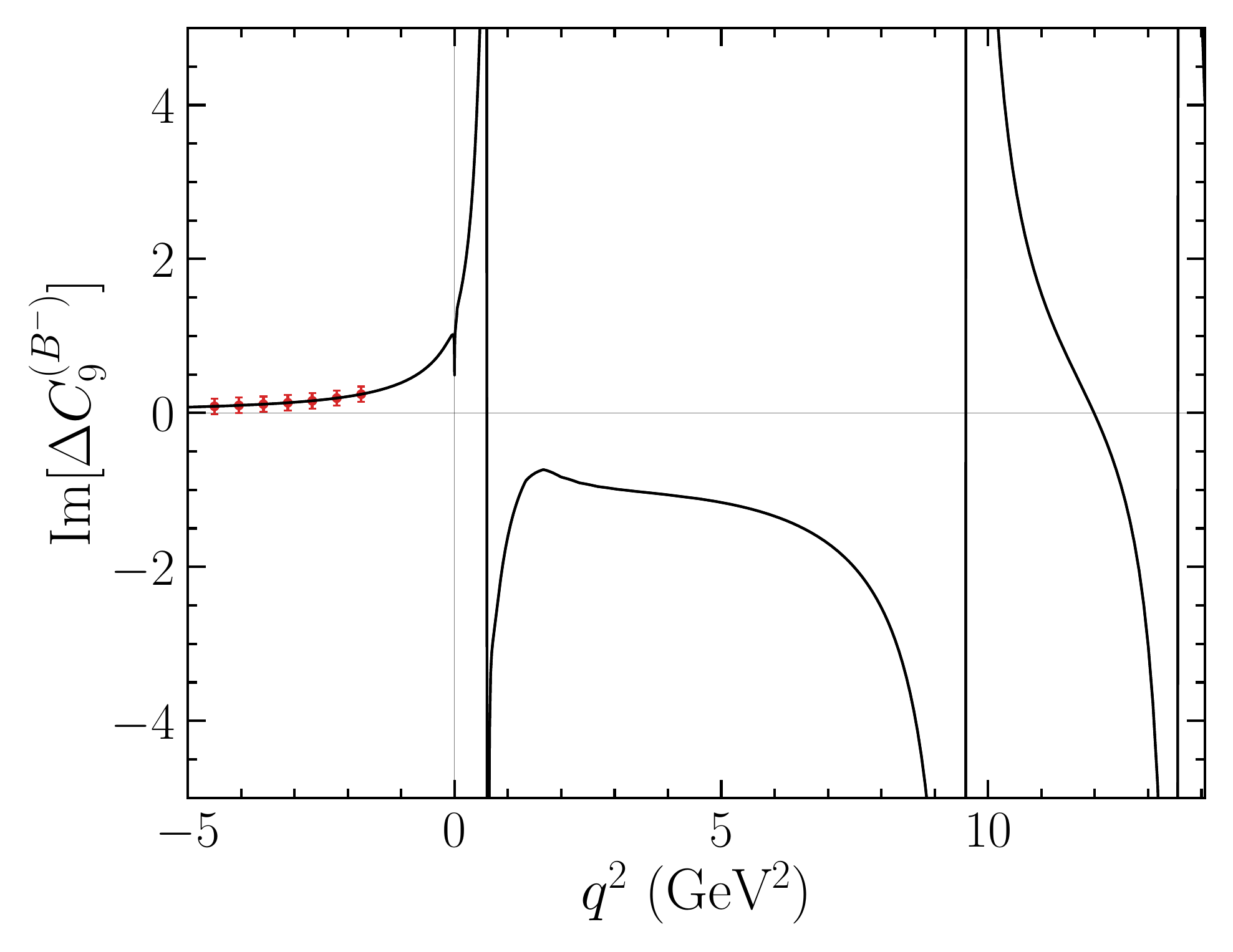}
        \includegraphics[width=0.48\textwidth,keepaspectratio]{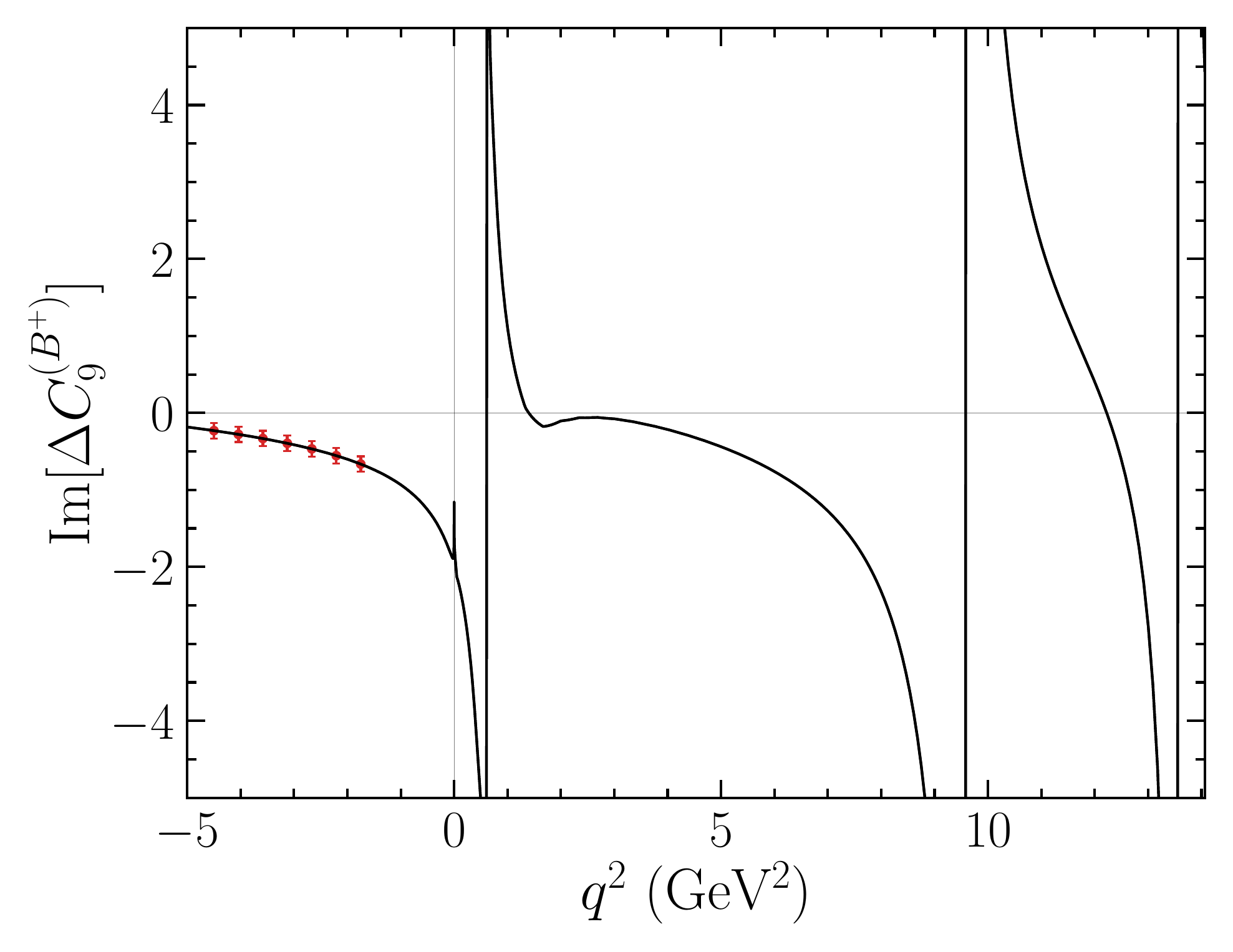}

        \caption{\label{fig:non_local_model}
            The model employed for the non-local contributions to  \BpmToPimm, along with the $q^2<0$ reference values and introduced in \autoref{sec:hadronic_input}.
        }
        \end{figure}

\begin{figure}[ht]
    \centering
    \includegraphics[width=0.48\textwidth,keepaspectratio]{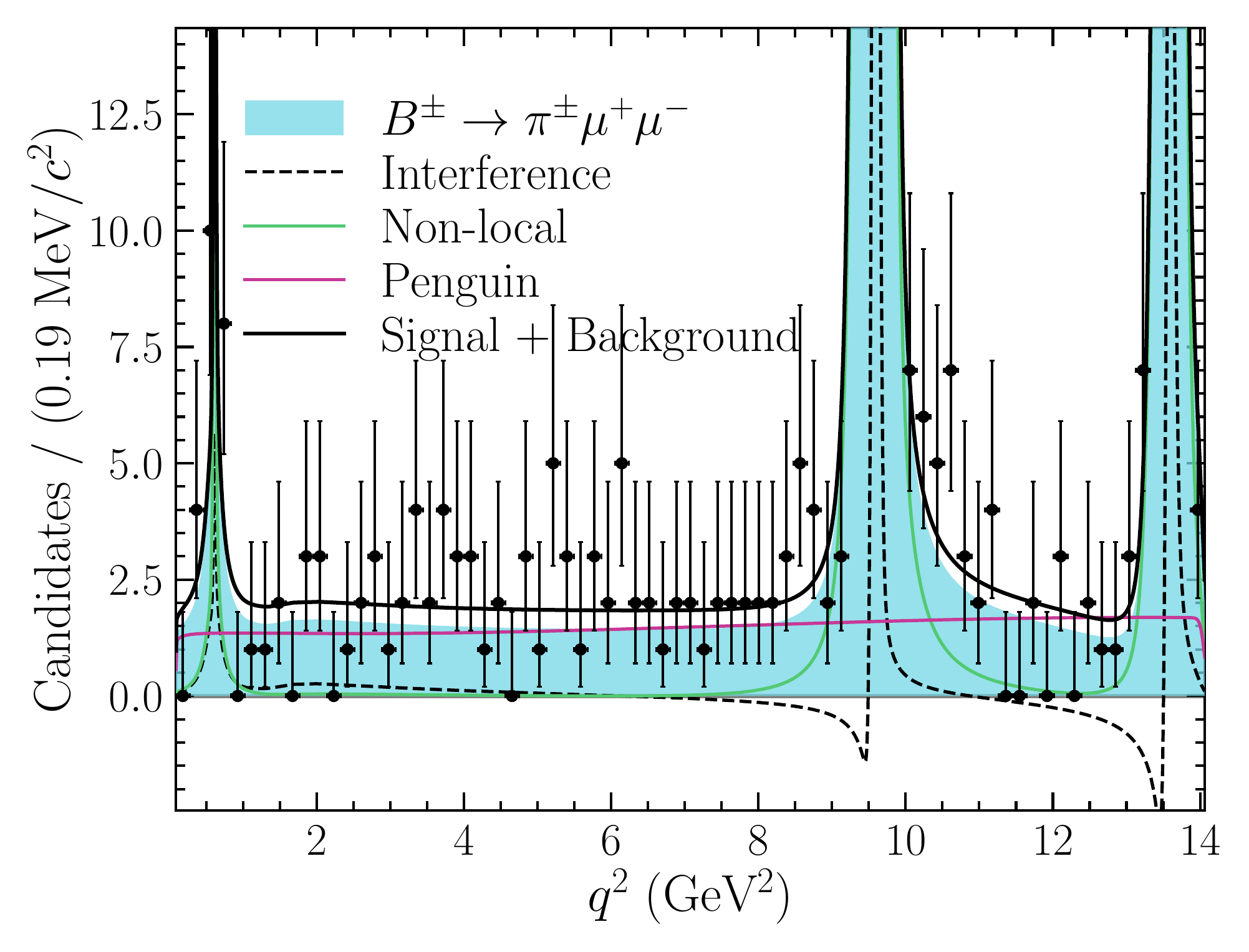}
    \includegraphics[width=0.48\textwidth,keepaspectratio]{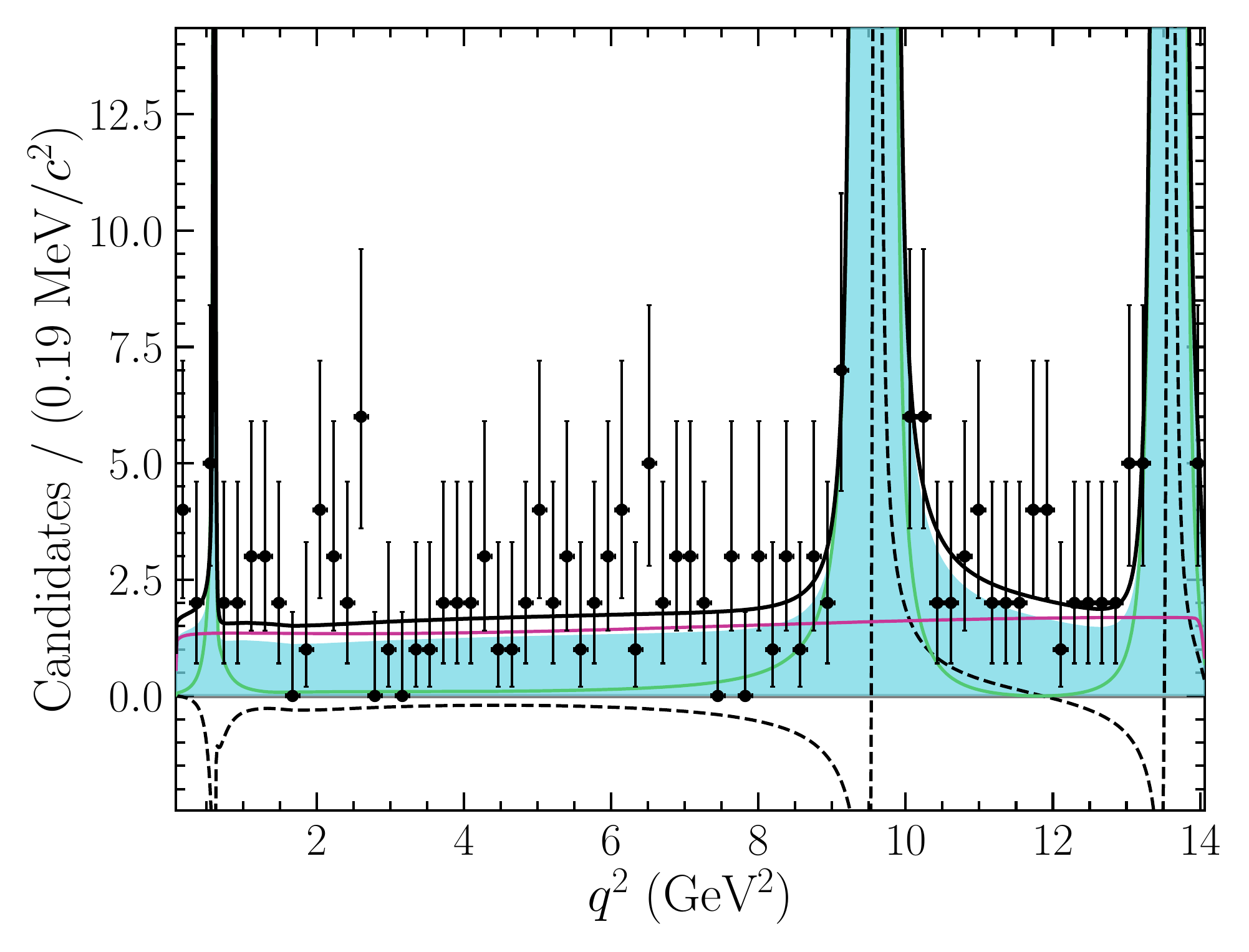}

    \includegraphics[width=0.48\textwidth,keepaspectratio]{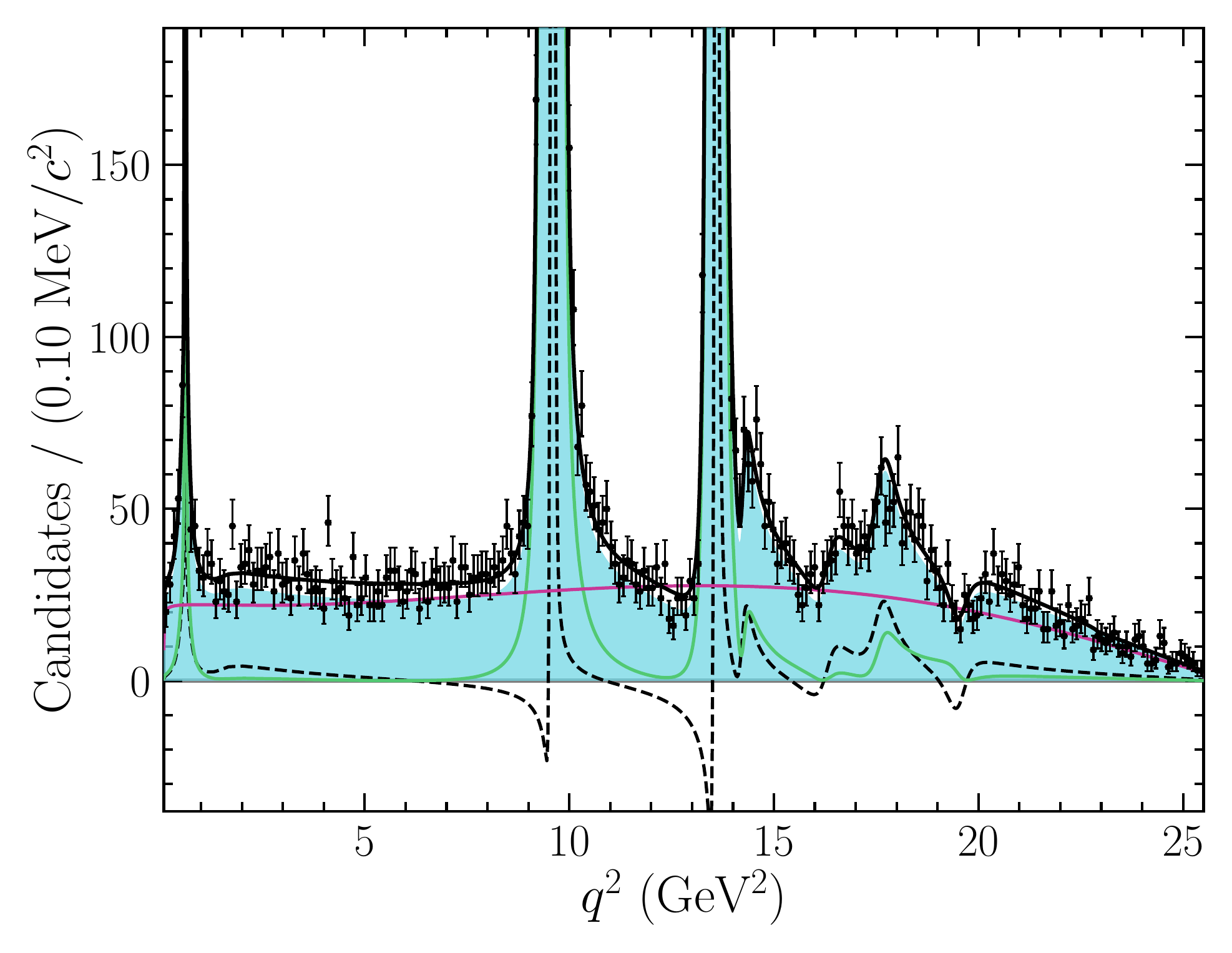}
    \includegraphics[width=0.48\textwidth,keepaspectratio]{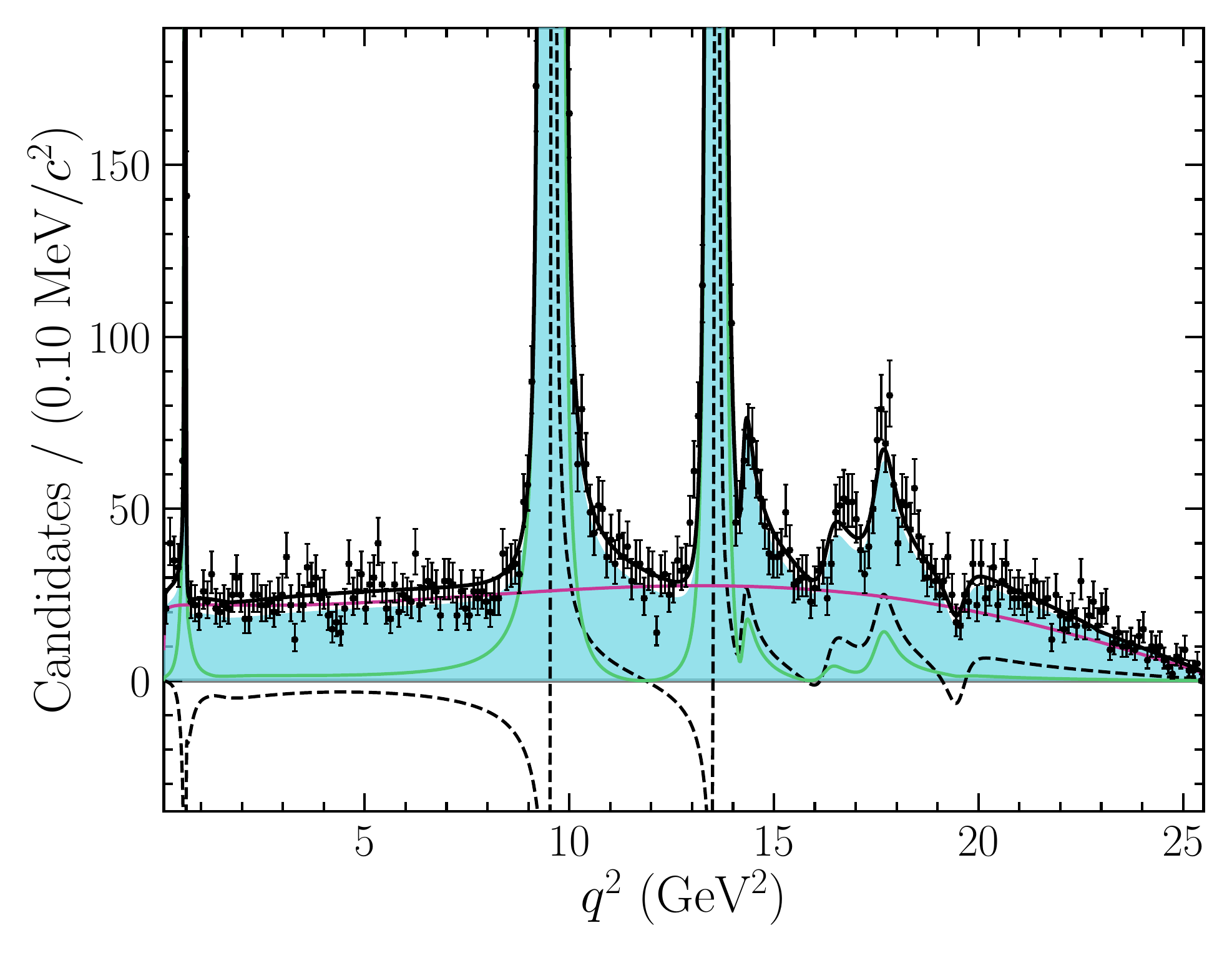}

    \caption{\label{fig:9fb}
        Pseudo-data generated to represent a selected \BpmToPimm dataset obtained from Run1+2 of LHCb data (top) and 300~fb$^{-1}$ (bottom), $B^+\rightarrow\pi^+\mu^+\mu^-$ (left) and $B^-\rightarrow\pi^-\mu^+\mu^-$ (right).
    }
\end{figure}

\section{Analysis Setup}
\label{sec:Fit_Description}

    We generate pseudo-datasets using the decay rate in \autoref{eq1} and keep the Wilson coefficients set to their SM values. The parameters that describe the local form factors are assigned to the central values obtained in Ref.~\cite{Leljak:2021vte}. The parameters in the description of the non-local form factors are instead obtained from a $\chi^2$ fit of our model described in \autoref{sec:hadronic} to the theoretical pseudo data points at $q^2<0$ computed in Ref.~\cite{Hambrock:2015wka}. Our results for the latter parameters are compatible with those of Ref.~\cite{Hambrock:2015wka}. We present our results for the non-local contributions expressed in terms of the quantity $\Delta C_9$ in \autoref{fig:non_local_model}.

    To ascertain a realistic expected precision on the parameters of interest from the fit to the \qsq spectrum of \BpmToPimm decays, we need to take into account the expected experimental $\qsq$ resolution $R(\qsq_{\textrm{reco}},\qsq)$ and the reconstruction efficiency $\varepsilon(\qsq)$. We use the experimental $\qsq$ resolution used in the LHCb analysis of \BpmToKmm decays in Ref.~\cite{LHCb:2016due}. Our choice is motivated by the expectation that this resolution is close, if not identical, to the LHCb resolution for \BpmToPimm decays. For the reconstruction efficiency, we take the \qsq shape of the efficiency reported in Ref.~\cite{LHCb:2016due} and extrapolate it linearly to the larger phase space of \BpmToPimm decays.  The final signal \qsq model is given by the convolution
    \begin{equation}
    \label{eq:reconstructed_rate}
    R(\qsq_{\textrm{reco}},\qsq)\otimes\left(\frac{d\Gamma}{d\qsq}\varepsilon(\qsq)\right),
    \end{equation}
    which is obtained through a fast Fourier transform.

    \noindent The signal yield is obtained using the expression
    \begin{equation}
        \displaystyle N_\mathrm{sig}^{B^\pm} = \mathcal{L}\, \alpha\, N_{B^\pm \to J/\psi K^\pm}\abs[\Big]{\frac{V_{cd}^{\phantom{*}}}{V_{cs}^{\phantom{*}}}}^2\frac{\int\frac{d\Gamma(B^\pm)}{d\qsq} d\qsq}{ \int|Y^{B^\pm}_{J/\psi}(\qsq)|^{2}d\qsq}, 
    \end{equation}
    where $\alpha$ is a factor that represents all relative efficiency effects such that the calculated signal yield is compatible with that of the measured yields in different $q^2$ bins in Ref.~\cite{LHCb:2015hsa}. The factor $N_{B^\pm \to J/\psi K^\pm}$ is the reconstructed yield of $B^\pm \to J/\psi(\to\mu^+\mu^-) K^\pm$ candidates taken from Ref.~\cite{LHCb:2016due} that used 3~fb$^{-1}$ of LHCb Run1 data. The ratio of CKM matrix elements converts $N_{B^\pm \to J/\psi K^\pm}$ to the number of expected $B^\pm \to J/\psi K^\pm$ decays. The factor $\mathcal{L}$ scales up the yields from 3~fb$^{-1}$ of LHCb Run1 data to future projections of LHCb integrated luminosities, including the increase in the $B$-hadron production cross-sections coming from the centre-of-mass energy changes of the LHC.  The factor $\frac{\int\frac{d\Gamma}{d\qsq} d\qsq}{ \int|Y^{\pm}_{J/\psi}(\qsq)|^{2}d\qsq}$ uses our model of the $B^\pm\to\pi^\pm\mu^+\mu^-$ decay rate to transform the $B^\pm\to J/\psi\pi^\pm$ yield into one across the entire \qsq phase space. Finally, the signal purity is estimated from Fig.~3 of Ref.~\cite{LHCb:2015hsa}, and the \qsq model of the background is taken from Fig.~3 of Ref.~\cite{LHCb:2016due} and modelled using a kernel density estimator. Pseudo-datasets are generated with a sample size corresponding to that expected in the current LHCb Run1+2 dataset (9~fb$^{-1}$) and future LHCb Upgrades of 23~fb$^{-1}$, 45~fb$^{-1}$ and 300~fb$^{-1}$.

    We perform unbinned maximum likelihood fits to these pseudo-datasets where the magnitude parameters $\eta^{B^\pm}_{J/\psi}$ and the parameters of the local form factors are fixed in the fit. Fixing these parameters incurs a systematic uncertainty, the size of which we assess in \autoref{sec:sensitivity}. In this fit configuration, we measure all the phases, including the phase of $C_9$, relative to that of $C_7$, which is fixed in the fit.

    Examples of pseudo-datasets representing 9~fb$^{-1}$ and 300~fb$^{-1}$ are presented in~\autoref{fig:9fb} along with the model employed in the pseudo-dataset generation. The non-local, penguin, and interference components of the model are shown separately.

    \subsection{Constraining the non-local contribution}
    \label{sec:hadronic_input}
    
        We relate the model of the non-local contribution $\Delta C_9^{B^\pm}$, as in \autoref{eq:dispersion}, to the sum of the various QCD factorisation and LCSR predictions at negative $q^2$ as in \autoref{eq:H_sum}. This relationship is visualised in \autoref{fig:non_local_model} where the red points denote the QCD factorisation and LCSR predictions, and the black line is our model of the non-local contributions to $B^\pm\rightarrow \pi^\pm\mu^+\mu^-$ decays.
    
        We can exploit this relation when fitting our model for the $B^\pm\rightarrow \pi^\pm\mu^+\mu^-$ decay rate, \autoref{eq1}, to data in the physical $q^2>0$ region through the introduction of a multivariate Gaussian factor to the likelihood function. This factor relates our dispersive non-local model to the theory reference values computed at different $q^2<0$ values, indicated by the red points in \autoref{fig:non_local_model}. The dimensionality of this multivariate Gaussian constraint is given by the number of negative $q^2$ points considered multiplied by four\footnote{For the real and imaginary components of the non-local amplitude for both $B^+$ and $B^-$.}. 
        
        The uncertainties of the reference values and the correlations between these uncertainties need to be taken into account in the multivariate constraint. As Ref.~\cite{Hambrock:2015wka} does not provide these correlations, we make the conservative assumption that all the uncertainties used to compute the theory terms are uncorrelated between \qsq points and between real/imaginary $B^+$/$B^-$ contributions. Assuming the uncertainties between the \qsq points are uncorrelated reduces the statistical power of the constraint.
        
        Constraints are placed on the magnitude parameters of the resonances $V=\rho(770)$, $\omega(782)$ and $\psi(2S)$ using measured central values and uncertainties for the \CP-averaged branching fractions of $\mathcal{B}(B\rightarrow V(\to \mu^+\mu^-)\pi)$ and of $A^V_{\CP}$~\cite{Workman:2022ynf}. These are essential for reliable fit convergence and are employed in all the fits discussed in this paper.

    \subsection{Choosing a $q^2$ range}
    \label{sec:q2_range}
    
        The region of \qsq above the open-charm threshold is particularly problematic due to the presence of multiple broad overlapping resonances that interfere with non-resonant contributions. With the number of signal decays expected in the existing LHCb Run1+2 dataset, it is unfeasible to float all the parameters associated with non-local contributions arising from open-charm states. Their impact, however, is sub-dominant for $q^2 \lesssim 14\,\mathrm{GeV}^2$. This leads us to fix these parameters and to restrict the phase space region for our analysis.

        We use the results from the $B^+\to K^+\mu^+\mu^-$ measurement of Ref.~\cite{LHCb:2016due} scaled by $|V_{cd}/V_{cs}|$ to fix the residues of the open-charm states. We further limit the phase space to $q^2_\mathrm{reco}<14.06~\mathrm{GeV}^2$.
        This cut is motivated by the fact that, taking into account resolution effects, contributions above the $\psi(3770)$ are negligible.
        
        In future datasets, such as those expected by LHCb's planned upgrade, the signal yield will be sufficient to fit the entire $q^2$ phase space with these non-local parameters floating. Therefore, the open charm region is included in the fits to 300~fb$^{-1}$ of pseudo-data, as presented in the bottom panels of \autoref{fig:9fb}.

    \subsection{Contamination from $B^\pm\to K^\pm\mu^+\mu^-$}
    
        The decay \BpmToKmm with a $K^\pm\to\pi^\pm$ misidentification is a potentially dangerous background to measurements of 
        \BpmToPimm as it is less CKM suppressed than the \BpmToPimm process. However, the binned measurement of the \BpmToPimm decay rate presented in Ref.~\cite{LHCb:2015hsa} demonstrated that the $B^\pm \to K^\pm\mu^+\mu^-$ background can be brought under control through the use of particle identification information from the ring-imaging Cherenkov (RICH) systems of LHCb. In this study, we assume the signal purity of a window of $\pm40$\;MeV around the $B^\pm$ mass as given in Ref.~\cite{LHCb:2015hsa}. However, the \BpmToPimm analysis of Ref.~\cite{LHCb:2015hsa} vetoed the regions associated with resonant dimuon contributions from $B^\pm\rightarrow K^\pm \psi(\rightarrow \mu^+\mu^-)$ decays, where $\psi$ is $J/\psi$ or $\psi(2S)$. Therefore, our assumed purity of \BpmToPimm decays in the \qsq regions near the large charmonia resonances is not valid. In principle, an experimental analysis that attempts to fit the entire \qsq spectrum of \BpmToPimm decays would have to adopt stricter particle identification criteria to reduce the background from $B^\pm\rightarrow K^\pm \psi(\rightarrow \mu^+\mu^-)$ decays down to a controllable level at the expense of signal efficiency. An experimental analysis may need to undertake some optimisation of the selection, including a background component for $B^\pm\rightarrow K^\pm \psi(\rightarrow \mu^+\mu^-)$ backgrounds in the fitted model and studying the impact on the signal precision. Therefore, dealing with this background is beyond the scope of our study.

\section{Experimental precision and prospects}
\label{sec:experimental_prospects_and_precision}

    To estimate the expected sensitivity to new physics and understand the impact of the $q^2<0$ constraint, we fit generated pseudo-datasets with and without the theoretical constraint at $q^2<0$ included in the likelihood. Each fit is initialised from multiple starting positions to avoid localised turning points in the likelihood space. The fit result with the largest likelihood is recorded.

\subsection{Fit stability}
    \label{sec:fit_stability}

        \begin{table}[t]
        \centering
        \begin{tabular}{@{}l l c c@{}}
        \toprule
        \multirow{2}{*}{Size of the dataset} & \multirow{2}{*}{Relative size} & \multicolumn{2}{c}{Fit success ($\%$)} \\
                                             &                                & w/o $q^2 < 0$ & with $q^2 < 0$ \\
        \midrule
            $9~\mathrm{fb}^{-1}$   & 1   & $36$  & $78$ \\  
            $23~\mathrm{fb}^{-1}$  & 2.5 & $83$  & $94$ \\ 
            $45~\mathrm{fb}^{-1}$  & 5   & $91$  & $95$ \\ 
            $300~\mathrm{fb}^{-1}$ & 33  & $100$ & $100$\\
        \bottomrule
        \end{tabular}
        \caption{\label{table:fit_stability_table}
            Stability of the fits to pseudo-data.
            The last column separates fits that do not use theoretical inputs at negative $q^2$ from those that do.
        }
        \end{table}
        
        With the signal yields expected from the 9~fb$^{-1}$ LHCb Run1+2 dataset, we find that the best-fit point of a significant fraction of pseudo-datasets lies in an unphysical region. The decay rate of \autoref{eq1} is not differentiable with respect to $C_{10}$ in the point $C_{10}=0$ due to the $|C_{10}|^2$ dependence in the decay rate.  As our likelihood minimisation relies on gradient descent methods, the algorithm fails when the estimated value of $C_{10}\approx0$. Reparametrising the likelihood in terms of $|C_{10}|^2$ (rather than $C_{10}$), we find the fits to these pseudo-datasets converge with negative $|C_{10}|^2$ values, implying an unphysical value for $|C_{10}|$. We, therefore, classify these 
        fits as failed and remove them from our ensembles of pseudo-experiments. We report the fraction of successful fits as a function of dataset size for fits with and without the $q^2<0$ constraint applied in \autoref{table:fit_stability_table}.

        We observe that the success rate of the fits increases by increasing the dataset size or by including the $q^2<0$ constraint in the likelihood. For smaller-sized datasets, where the fraction of successful fits is low, imposing some additional assumption on the new physics model, for example, $C_{10}^\mathrm{NP} = - C_{9}^\mathrm{NP}$ (where $C_i^\mathrm{NP} = C_i - C_i^\mathrm{SM}$), improves fit stability at the expense of introducing a model dependence.

\subsection{Assessing sensitivity to new physics}

    \label{sec:sensitivity}
    
    \begin{figure}[t!]
    \centering

    \includegraphics[width=0.48\textwidth,keepaspectratio]{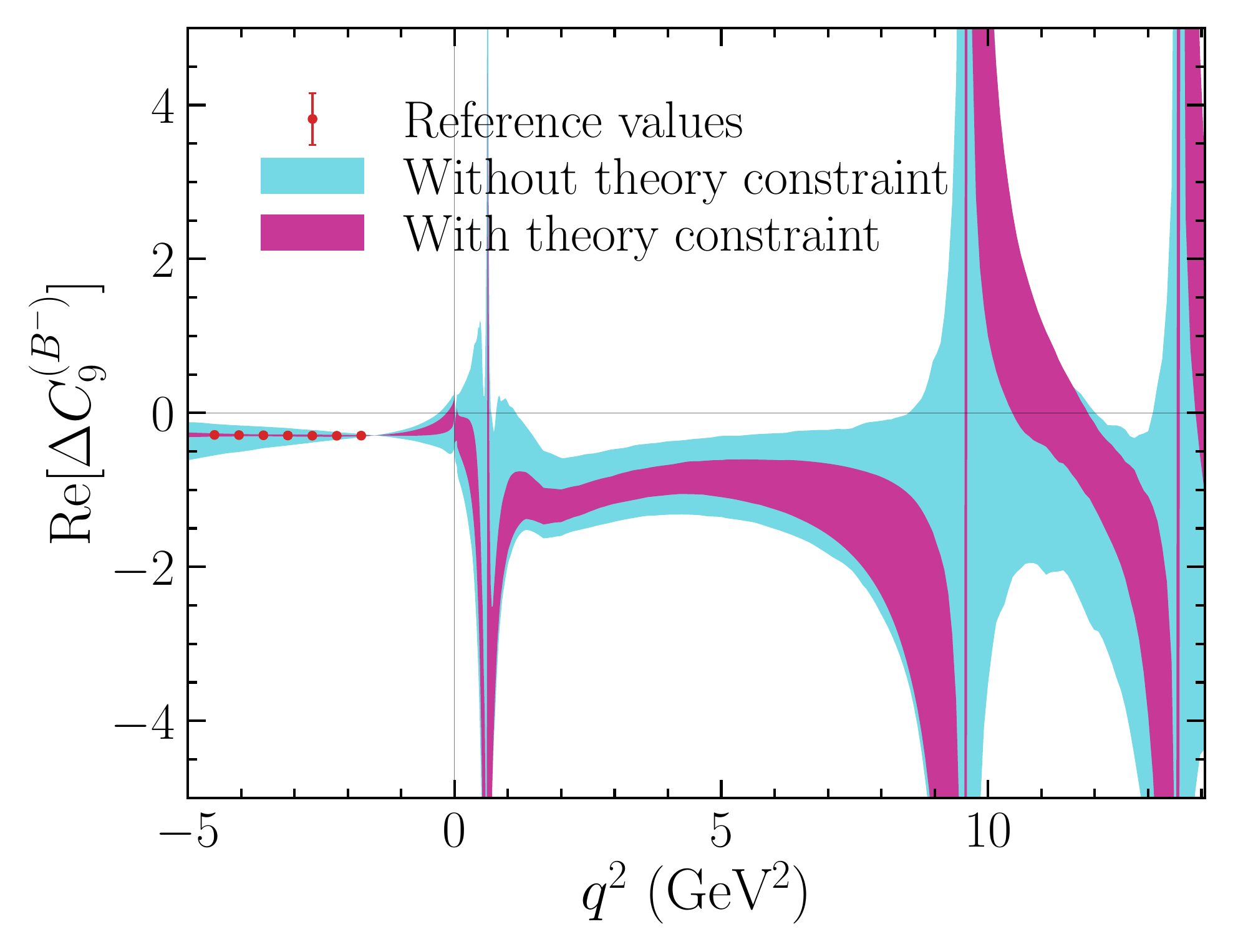}
    \includegraphics[width=0.48\textwidth,keepaspectratio]{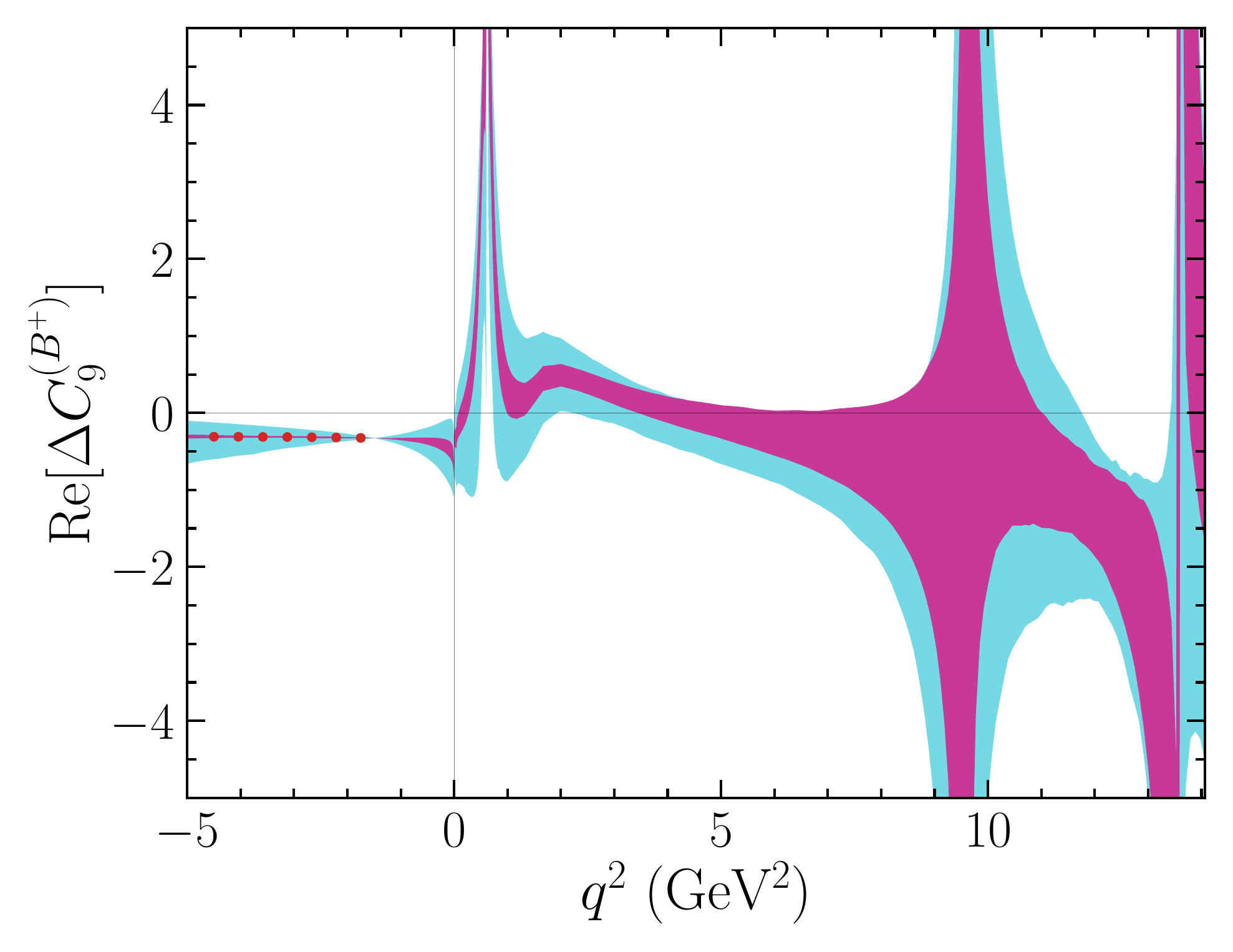}

    \includegraphics[width=0.48\textwidth,keepaspectratio]{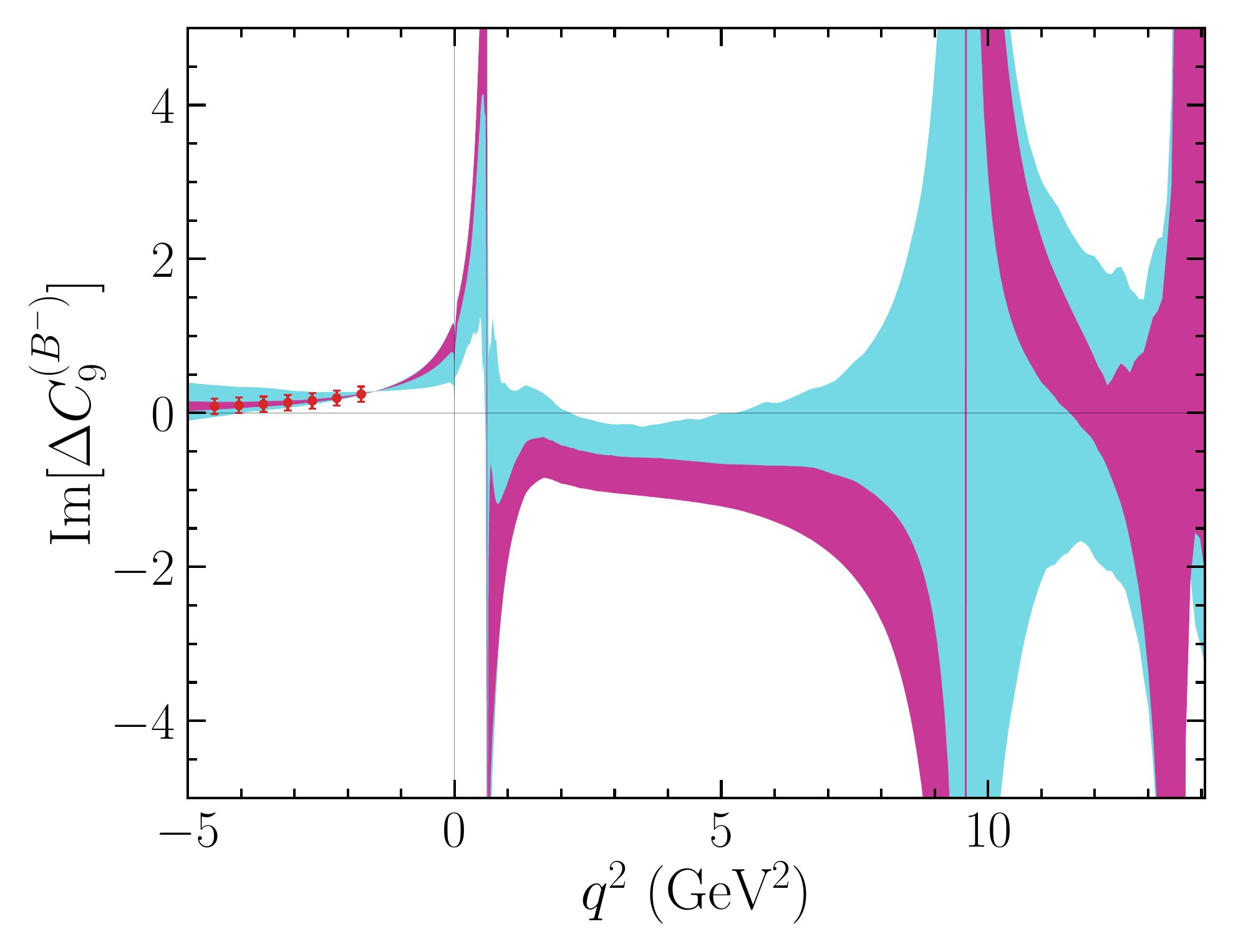}
    \includegraphics[width=0.48\textwidth,keepaspectratio]{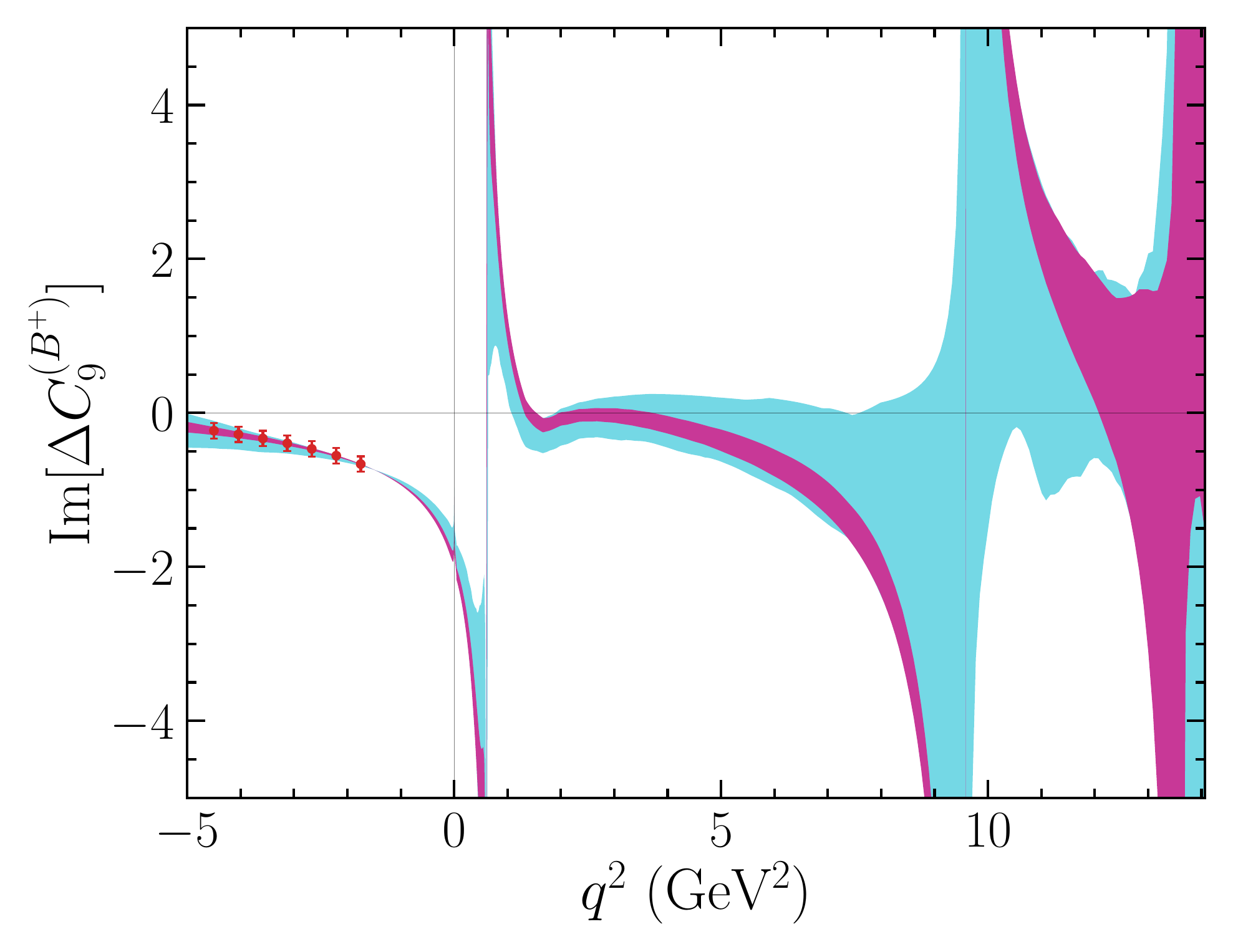}

    \includegraphics[width=0.48\textwidth,keepaspectratio]{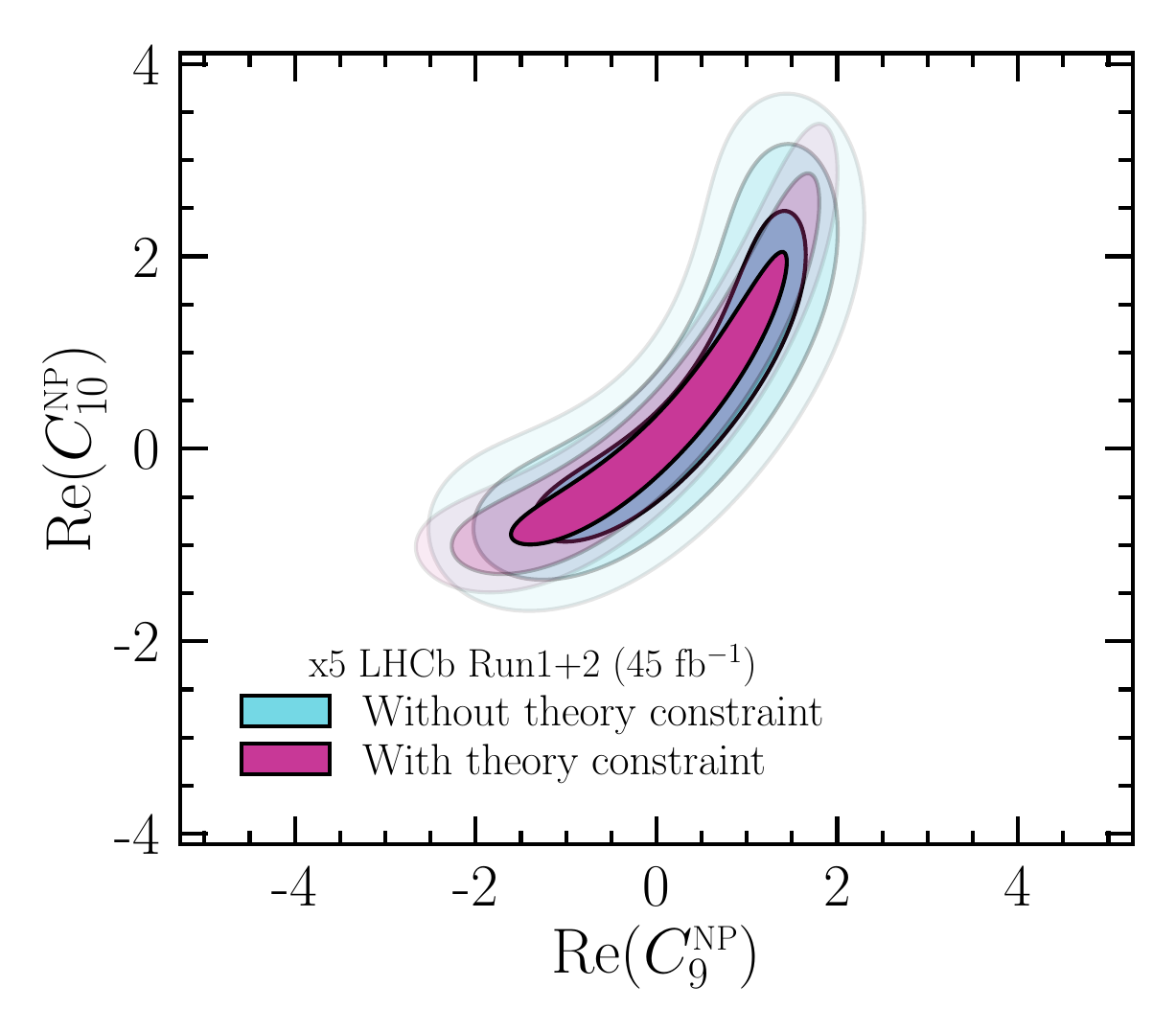}
    \includegraphics[width=0.48\textwidth,keepaspectratio]{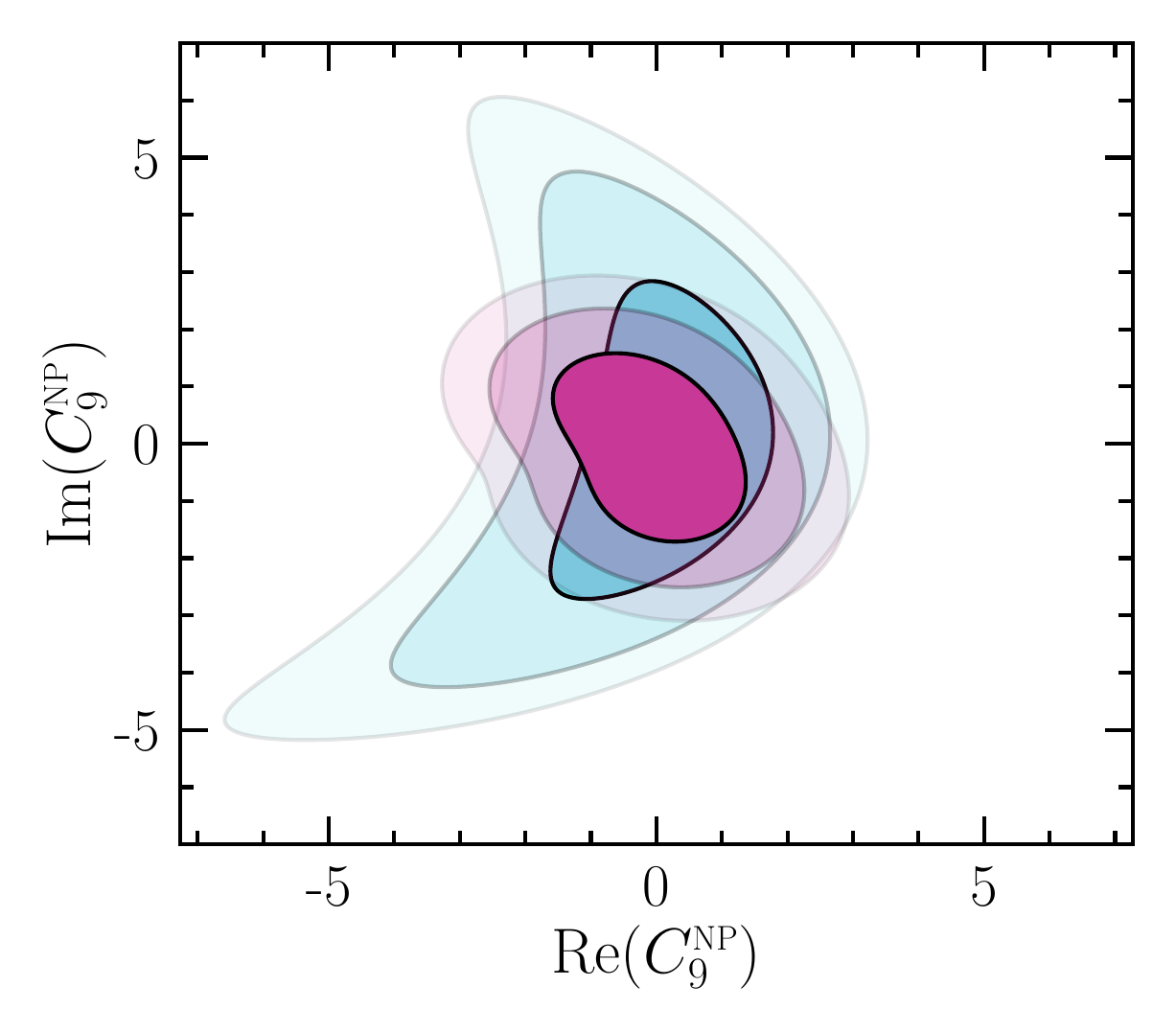}

    \caption{\label{fig:non_local_sens}
        The statistical sensitivity obtained from fits to pseudo-datasets representative of x5 the expected LHCb Run1+2 yields showing (top and middle) the $68\%$ intervals for the non-local component of $\Cnineeff(q^2)$; (bottom) the $68\%$, $95\%$ and $99\%$ intervals for the Wilson coefficients $\textrm{Re}(C_{10}^{\textrm{NP}})$, $\textrm{Re}(C_9^{\textrm{NP}})$ and $\textrm{Im}(C_9^{\textrm{NP}})$.
    }
    \end{figure}

    When employing the $q^2<0$ constraint, we observe a significant improvement in the statistical precision of the non-local contributions, as shown in \autoref{fig:non_local_sens}. This improvement subsequently translates into gains in the statistical precision of the new physics parameters $\textrm{Re}(C_{10}^{\mathrm{NP}})$, $\textrm{Re}(C_9^{\mathrm{NP}})$ and $\textrm{Im}(C_9^{\mathrm{NP}})$. The phases of the resonances and the parameters describing the $Y^{B^\pm}_{\text{2P}, c\bar{c}}(q^2)$ contributions exhibit significantly reduced uncertainties when employing the $q^2<0$ constraint. In contrast, the gains in precision to the magnitude parameters of the resonances are modest as the sensitivity to these parameters is dominated by the prior knowledge of their branching fractions, as mentioned in \autoref{sec:hadronic_input}. Taking the best-fit points of an ensemble of pseudo-experiments, we construct confidence intervals that illustrate the estimated sensitivity to the short-distance parameters $\textrm{Re}(C_{10}^{\mathrm{NP}})$, $\textrm{Re}(C_9^{\mathrm{NP}})$ and $\textrm{Im}(C_9^{\mathrm{NP}})$. These intervals are presented in the lower panels of \autoref{fig:non_local_sens} for fits to pseudo-datasets representing 45~fb$^{-1}$ of LHCb data both with and without the $q^2<0$ theory constraint. 

    \begin{figure}[t!]
    \centering

    \includegraphics[width=0.48\textwidth,keepaspectratio]{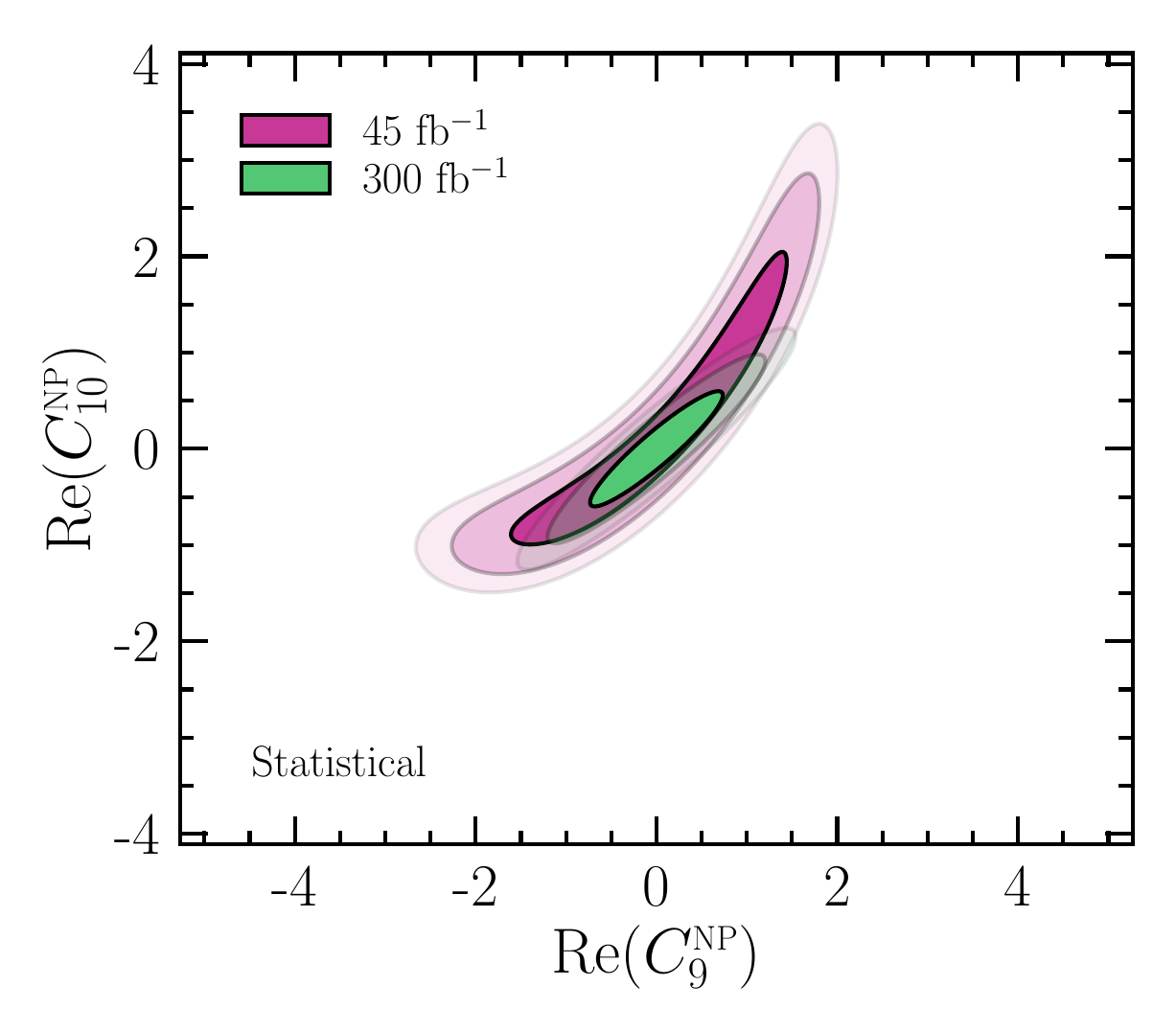}
    \includegraphics[width=0.48\textwidth,keepaspectratio]{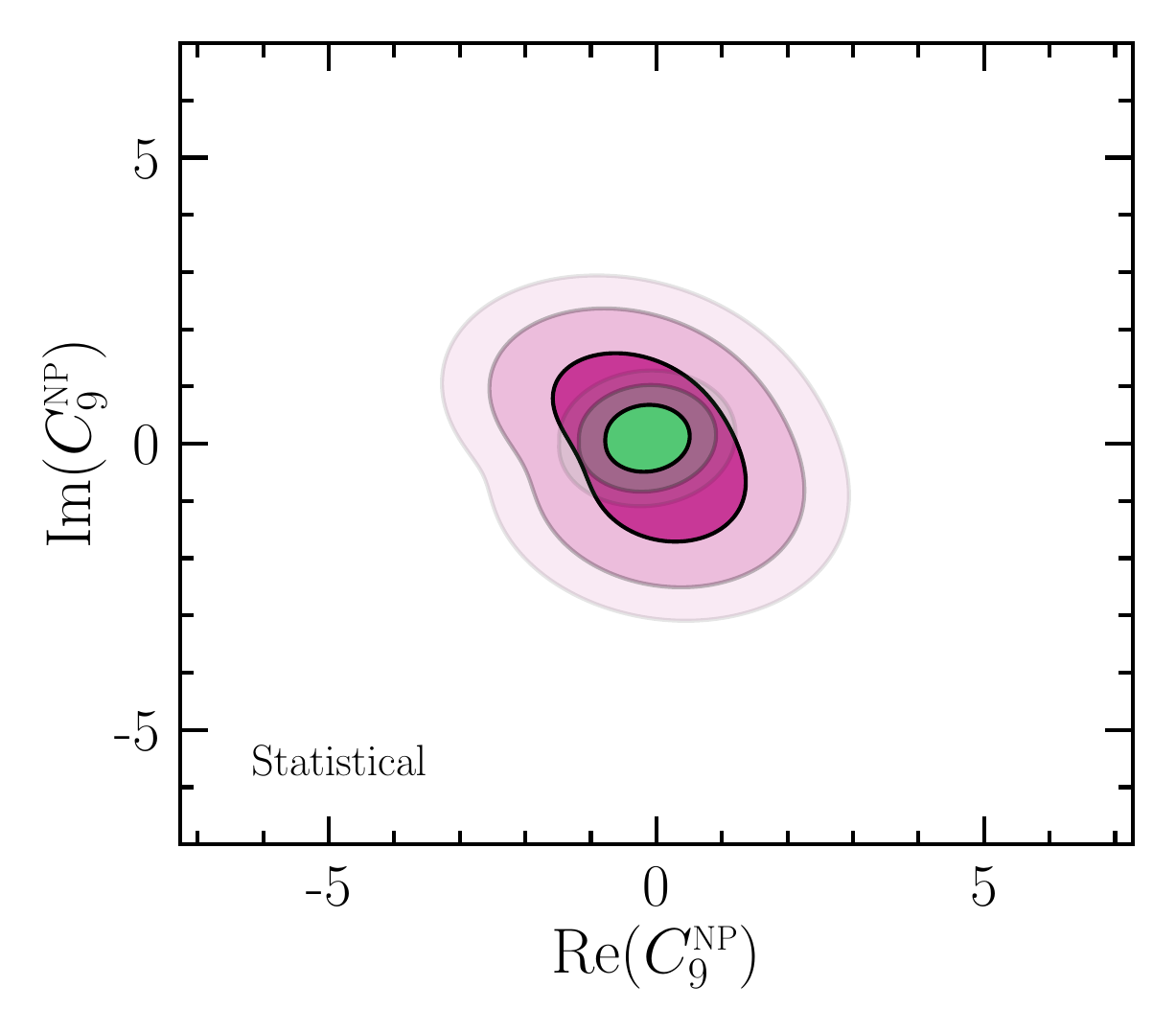}
    
    \includegraphics[width=0.48\textwidth,keepaspectratio]{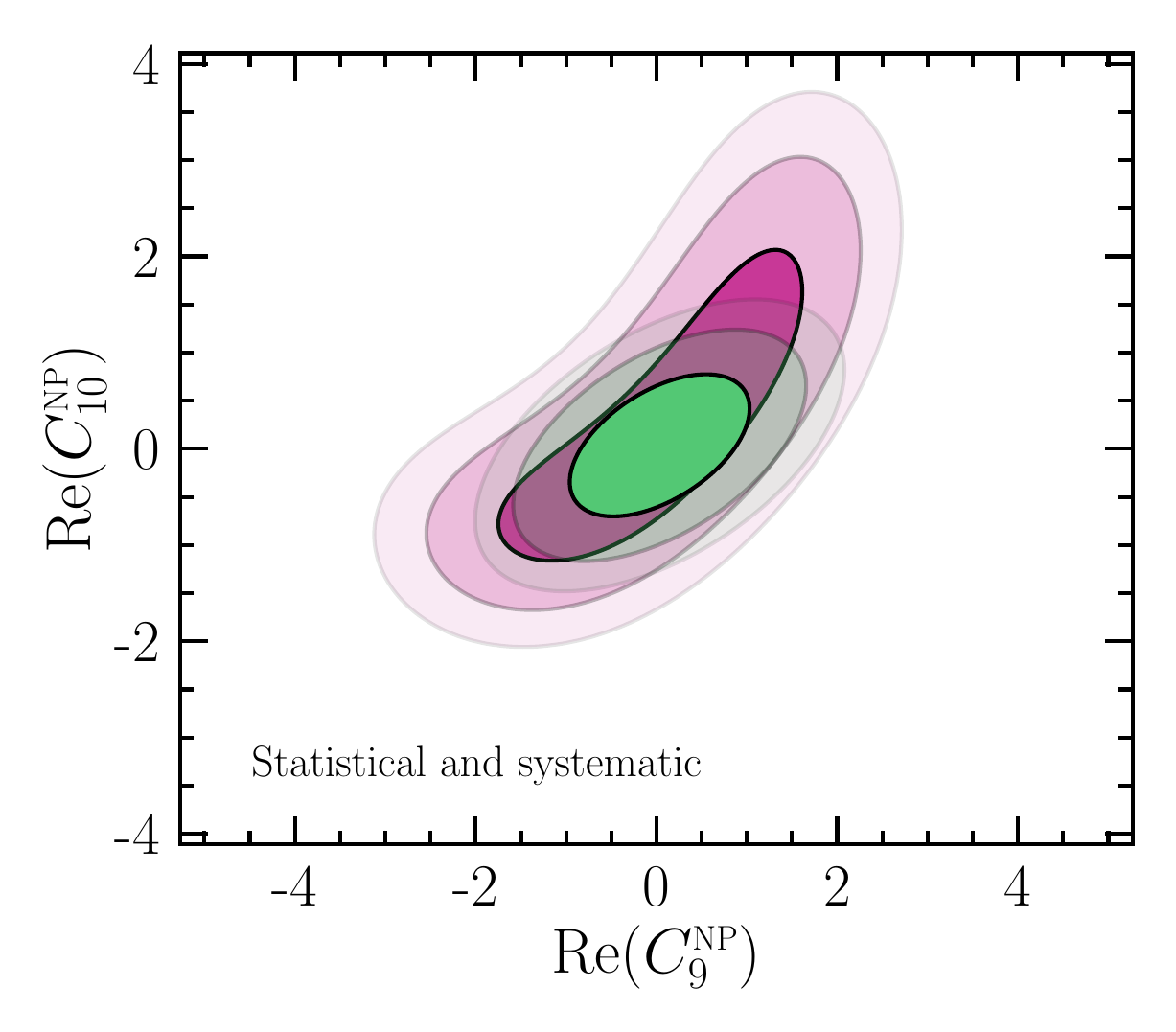}
    \includegraphics[width=0.48\textwidth,keepaspectratio]{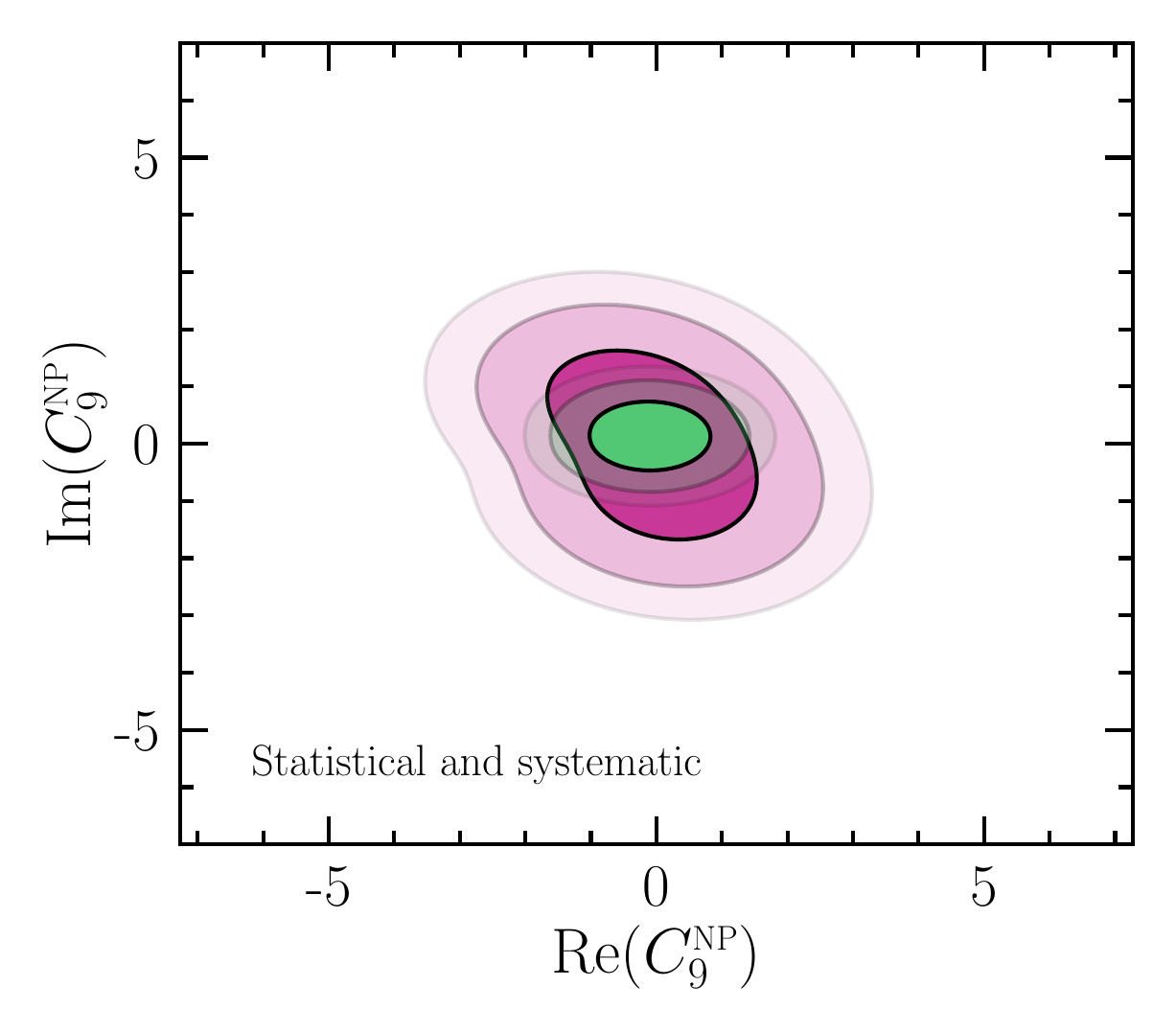}

    \includegraphics[width=0.48\textwidth,keepaspectratio]{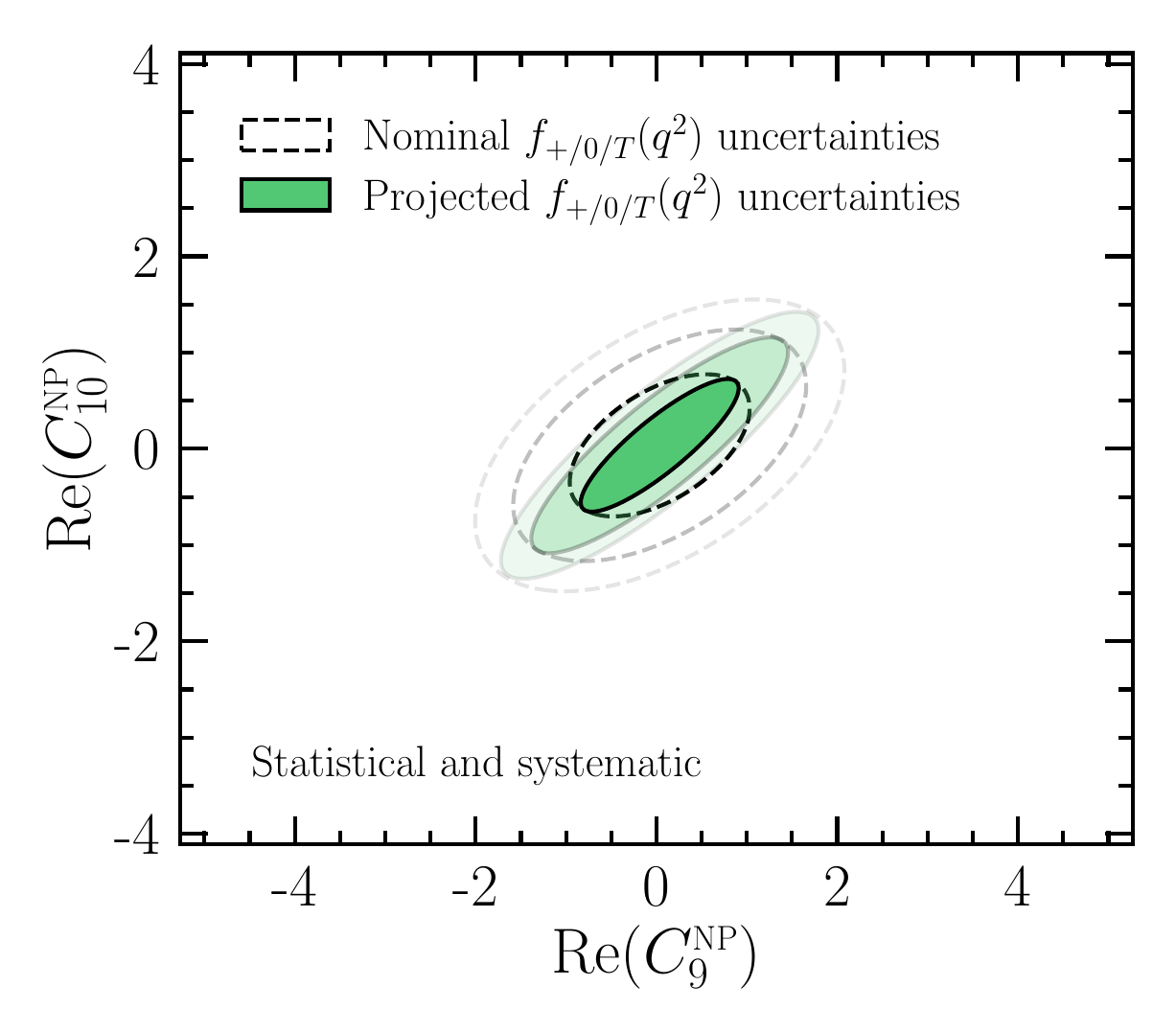}
    \includegraphics[width=0.48\textwidth,keepaspectratio]{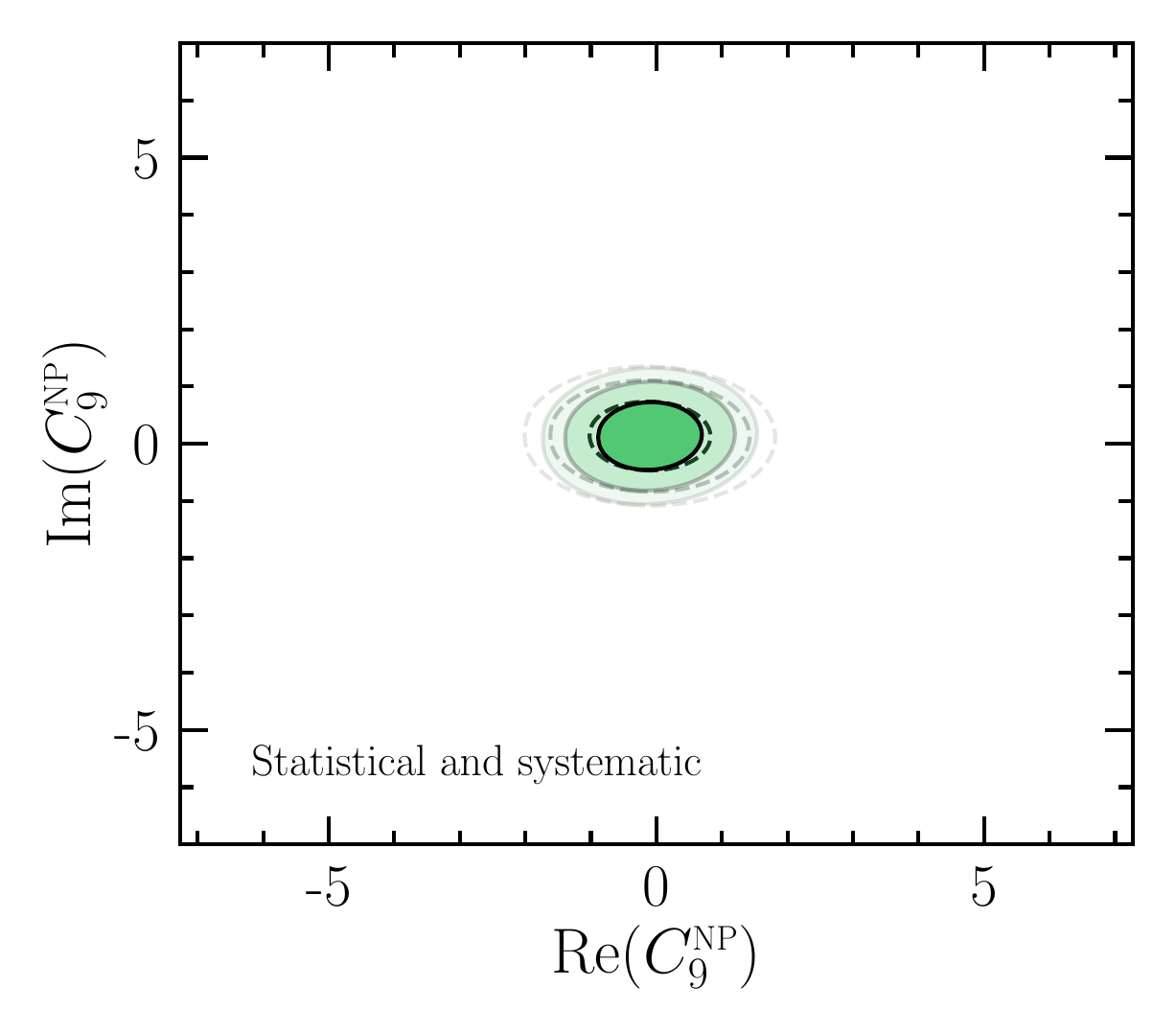}
    
    \caption{\label{fig:banana_plots}
        Two-dimensional intervals on planes of $\textrm{Re}(C_9^{\textrm{NP}})$ versus $\textrm{Re}(C_{10}^{\textrm{NP}})$ and $\textrm{Re}(C_9^{\textrm{NP}})$ versus $\textrm{Im}(C_9^{\textrm{NP}})$ showing (top) statistical uncertainty-only intervals (middle) intervals with the inclusion of systematic uncertainties (bottom) a comparison of intervals for 45~fb$^{-1}$ with current local form factor uncertainties (dashed) and with a projected improvement (solid).
    }
    \end{figure}

    The systematic uncertainties that arise from fixing $\eta^{B^\pm}_{J/\psi}$ and the local form factor parameters in the fit are computed and folded into the statistical confidence intervals. These systematic uncertainties are obtained using SM pseudo-experiments for  45~fb$^{-1}$ and 300~fb$^{-1}$ scenarios separately. This is done for two reasons. Firstly, the 300~fb$^{-1}$ fits employ the entire $q^2$ region and float more parameters of the non-local model, including those of the open-charm resonances. Secondly, the uncertainties of the $B^\pm\rightarrow J/\psi(\rightarrow \mu^+\mu^-) \pi^\pm$ branching fractions are scaled for the 300~fb$^{-1}$ scenario according to a projected improvement in precision. The estimated improvement in precision is based on the following assumptions: we assume no improvement in the \CP-averaged branching fraction measurement of $B^+\rightarrow J/\psi \pi^\pm$ decays, as it is already systematically limited~\cite{BELLE:2019xld}; we scale the statistical uncertainty of the statistically limited $A_{\CP}$ measurement in $B^+\rightarrow J/\psi \pi^\pm$ decays~\cite{LHCb:2016ehk} by the expected gain in signal yields at the LHCb experiment; and finally, we assume no improvement in the uncertainty of $\mathcal{B}(J/\psi\rightarrow \mu^+\mu^-)$ as it is also systematically dominated~\cite{BESIII:2013csc}. Intervals for the Wilson coefficients $\textrm{Re}(C_{10}^\mathrm{NP})$, $\textrm{Re}(C_9^\mathrm{NP})$ and $\textrm{Im}(C_9^\mathrm{NP})$ both with and without these systematic uncertainties are presented in \autoref{fig:banana_plots}. These intervals represent the expected sensitivity to these parameters when including the $q^2<0$ constraint and are presented for both the 45~fb$^{-1}$ and 300~fb$^{-1}$ scenarios.
    
    The uncertainties of the local form-factor coefficients are the primary source of systematic uncertainty on all the short-distance parameters: $\textrm{Re}(C_{10}^\mathrm{NP})$, $\textrm{Re}(C_9^\mathrm{NP})$ and $\textrm{Im}(C_9^\mathrm{NP})$. We, therefore, stress the importance of reducing form-factor uncertainties alongside the coming increase of signal yield expected from future runs of the LHC. We present an illustrative example to highlight this point. We overlay intervals obtained using smaller form factor uncertainties in the lower panels of \autoref{fig:banana_plots}. Here, we assume improved calculations could produce uncertainties three times smaller. This would be in line with the improvements achieved for $B\to K^{(*)}$ in Ref.~\cite{Gubernari:2023puw}. The improvement in the intervals is significant and brings the result much closer to the statistical-only intervals in the top panels of \autoref{fig:banana_plots}. 

    Given that the flavour anomalies could be indicating the presence of large lepton flavour universality violating (LFUV) contributions to $C_9^{\tau}$, the study of $C_9^{\tau}$ through $b\to d\lbrace e^+e^-,\mu^+\mu^-\rbrace$ transitions is an increasingly interesting subject~\cite{Bobeth:2011st, Glashow:2014iga, Alonso:2015sja, Barbieri:2015yvd, Crivellin:2017zlb, Buttazzo:2017ixm, Capdevila:2017iqn, Bordone:2018nbg}. As demonstrated in Ref.~\cite{Cornella:2020aoq}, large non-local contributions from $C_9^{\tau}$ can be imprinted into the \qsq spectrum of $B^\pm\to K^\pm\mu^+\mu^-$ decays. Larger futures datasets of \BpmToPimm decays could be used to study the \qsq distribution of \BpmToPimm decays by including a $C_9^{\tau}$ contribution for $\tau\tau$ re-scattering to $\mu\mu$. Additionally, with larger datasets, it would be possible to lift the model-dependence of the open charm continuum model by floating individual components of the $Y^{B^\pm}_{\text{2P}, c\bar{c}}(q^2)$ model or by allowing for \CP-violation. Increasing the complexity of the non-local model will only increase the relevance of the $q^2<0$ constraint. Finally, in the future, it will be possible to fit the \BpmToPimm decay rate for the presence of new physics with scalar and tensor Wilson coefficients. This would require a 2D fit of $q^2$ and the lepton helicity angle $\cos(\theta_\ell)$ using the double-differential decay rate~\cite{Bobeth:2007dw,Bobeth:2012vn}. Employing the $q^2<0$ information will be essential to maximise sensitivity to new physics in all these studies.

\section{Summary and Conclusions}
\label{sec:conclusion}

    This paper presents an approach that maximises the sensitivity of new physics searches in \BpmToPimm transitions. 
    We employ a dispersive model to perform unbinned maximum likelihood fits to both the measured dimuon \qsq spectrum of \BpmToPimm decays and to theoretical constraints on the non-local contributions at $q^2<0$. Our approach ensures that the size and the \qsq dependence of non-local contributions to \BpmToPimm transitions in the $q^2<0$ region align with predictions. We perform fits to pseudo-datasets and demonstrate the expected sensitivity to \CP-violating and \CP-conserving contributions for a variety of upcoming datasets. We observe that including the theoretical constraints markedly increases the fit stability and improves the sensitivity to non-local parameters and, consequently, to the Wilson coefficients. Variations in the modelling of the non-local amplitude above the open-charm threshold were found to have a negligible impact on the extracted values of the Wilson coefficients compared to their statistical precision. We conclude that without increased model dependence, an unbinned analysis of the Run1+2 LHCb dataset would be challenging due to poor fit stability. Instead, we present the expected sensitivity for the future scenarios of 45~fb$^{-1}$ and 300~fb$^{-1}$ of LHCb data. 
    We include systematic effects arising from our incomplete knowledge of the $B^\pm\to J/\psi(\to\mu^+\mu^-)\pi^\pm$ branching fractions and local form-factors. We find that uncertainties due to the local form factor knowledge currently form the dominant systematic uncertainty. This highlights that improving the precision of local form factors will be an essential step to fully exploit the physics potential of future datasets.
    
\bigskip

\section*{Acknowledgements}

    A.M.M.~acknowledges support by the UK Science and Technology Facilities Council (grant number ST/W000490/1). D.v.D.~acknowledges support by the UK Science and Technology Facilities Council (grant numbers ST/V003941/1 and ST/X003167/1).

\appendix

\bibliographystyle{JHEP}
\bibliography{main}

\end{document}